\documentclass[aps, prd, 10pt, notitlepage, superscriptaddress, nofootinbib,numbers,showpacs]{revtex4-1}

\usepackage[utf8]{inputenc}
\usepackage[T1]{fontenc}
\usepackage{anyfontsize}
\usepackage{amsmath}
\usepackage{amssymb}
\usepackage{amsfonts}
\usepackage{mathrsfs}
\usepackage{enumitem}

\usepackage{microtype}
\usepackage[normalem]{ulem}

\usepackage{graphicx}
\usepackage{epsfig}
\usepackage{float}

\usepackage[dvipsnames]{xcolor}
\usepackage{hyperref}
\hypersetup{
    colorlinks=true,
    citecolor=Purple,
    linkcolor=Purple,
    urlcolor=Purple,
    linktocpage=true,
    breaklinks=true
}
\usepackage{soul} 
\usepackage[capitalize]{cleveref}

\begin{document}
\title{Quantum improved wormholes in the Dekel-Zhao dark matter halo}

\author{Jonathan A. Rebouças}
\email{jalvesreboucas@ifce.edu.br}
\affiliation{Instituto Federal de Educação Ciências e Tecnologia do Ceará (IFCE), Iguatu, Brazil}
\affiliation{Universidade Estadual do Cear\'a (UECE), Faculdade de Educa\c{c}\~ao, Ci\^encias e Letras de Iguatu, Av. D\'ario Rabelo s/n, Iguatu - CE, 63.500-00 - Brazil.}

\author{Celio R. Muniz}
\email{celio.muniz@uece.br}
\affiliation{Universidade Estadual do Cear\'a (UECE), Faculdade de Educa\c{c}\~ao, Ci\^encias e Letras de Iguatu, Av. D\'ario Rabelo s/n, Iguatu - CE, 63.500-00 - Brazil.}

\author{Francisco Bento Lustosa}
\email{chico.lustosa@uece.br}
\affiliation{Universidade Estadual do Cear\'a (UECE), Faculdade de Educa\c{c}\~ao, Ci\^encias e Letras de Iguatu, Av. D\'ario Rabelo s/n, Iguatu - CE, 63.500-00 - Brazil.}

\author{Edson Otoniel }
\email{cedson.otoniel@ufca.edu.br}
\affiliation{Universidade Federal do Cariri (UFCA), Instituto de Forma\c{c}\~ao de Educadores - IFE,  R. Oleg\'ario Emidio de Araujo S/N, Brejo Santo - CE, 63.260-000 - Brazil}

\date{\today}
\begin{abstract}
\begin{center}
\textbf{Abstract}
\end{center}
This work presents and investigates novel traversable wormhole solutions within the framework of Asymptotically Safe Gravity (ASG), sourced by a dark matter halo modeled by the Dekel–Zhao density profile. The scale-dependent gravitational coupling $G(k)$, derived from the ASG renormalization group flow in the infrared regime, is incorporated directly into the field equations, providing a consistent description of quantum gravitational corrections even at astrophysical scales. The combined effects of the running coupling (parameterized by $\xi$) and the dark matter characteristics determine the geometric structure and physical viability of the wormhole. The solutions satisfy the flare-out and asymptotic flatness conditions within restricted parameter domains, exhibiting enhanced curvature near the throat due to ASG corrections. Null Energy Conditions are necessarily violated at the throat, and stability analysis based on the adiabatic sound speed as well as the modified Tolman–Oppenheimer–Volkoff equation reveal that quantum effects from ASG counteract the destabilizing influence of dark matter. Phenomenologically, the wormhole shadow radius increases nearly linearly with $\xi$, lying within the Event Horizon Telescope bounds for Sgr~A* when $\xi/M \simeq 0.8$--$0.9$, thus suggesting that ASG-corrected wormholes may represent observable signatures of quantum gravity in the strong-field regime.
\end{abstract}
\maketitle
\def\HMS{{\scriptscriptstyle{HMS}}}
\section{INTRODUCTION}
\label{S:intro}
The study of wormholes has interested physicists since the early days of General Relativity \cite{flamm1916beitrage}. From the seminal work of Einstein and Rosen \cite{EinsteinRosen1935} up to the fundamental contributions of Morris and Thorne \cite{Morris1988}, they remained largely a theoretical curiosity. Wormholes are solutions of Einsteins equations that would function as a tunnel between different ``sheets'' of spacetime (possibly connecting distant patches of the universe or even different neighboring universes). In \cite{Morris1988} it was shown that some form of ``exotic'' matter is required to stabilize these solutions to make them traversable wormholes (TWs). Dark energy or Casimir vacuum energy have been used as possible sources of this exotic matter in multiple wormhole solutions studied in the literature \cite{Lemos2003, Garattini2019} and modified gravity theories have been used to construct wormhole solutions without the need for exotic matter \cite{Capozziello:2022zoz}. If those geometric structures do exist in the universe their stability might be dependent on the quantum nature of the gravitational field, which has also been explored in the literature using different approaches like Loop Quantum Gravity \cite{Cruz_2024} or the Asymptotically Safe Gravity approach \cite{Alencar2021, Nilton2021, Nilton2022}. Beyond the motivation surrounding the possibility of spacetime travel using warp drives \cite{Lobo_2004}, in recent years there has also been a wider interest in the study of microscopical wormholes in the context of the ER=EPR conjecture \cite{Maldacena2013, Lobo2014}. Considering they could exist, one is faced with the challenge of deriving their properties and phenomenological characteristics in different settings. In the case of macroscopic astrophysical wormholes, one of the main difficulties will be how to differentiate this compact objects from ``ordinary'' black holes \cite{Bambi2014, Valentin2022}. Although the last decades have seen an astonishing number of confirmations of predictions from General Relativity and the Standard Model of Cosmology \cite{EventHorizonTelescope:2022wkp, LIGOScientific:2020ibl,Planck2018,Turener2022}, the golden age of precision cosmology and the continued advance of our astrophysical surveys has also allowed us to probe the limits of our theories and models to a point where they might require fundamental revision \cite{Berti_2015, Ashtekar2022, Vagnozzi_2023}. In cosmology, Dark Energy and Dark Matter continue to challenge our observations and experiments. Using a very detailed analysis of the large scale structure of our universe, the DESI collaboration has found evidence that the equation of state for the Dark Energy component might be variable ruling out its identification with a cosmological constant or a simple model of vacuum energy \cite{Adame_2024, Adame_2025, DESI2025}. Previously considered to be a matter of time, Dark Matter detection in particle accelerators continues to elude experimentalists \cite{Silk2010, cirelli2024darkmatter}. Both dark components could be explained by modified gravity theories, such as $f(R)$ or $f(R, T)$ \cite{Capozziello_2006, Harko2011}, and also could be connected to the fundamental nature of quantum gravity\footnote{Most models of modified gravity theories are motivated or at least inspired by grand unification proposals or quantum gravity arguments, however, not all quantum gravity theories lead to $f(R)$-like modifications of Einstein's General Relativity. Some examples of how modified gravity theories can be directly linked to quantum gravity approaches can be found in \cite{Olmo_2009, Lobato2019}.}. Black holes are also a fundamental part of the discussion regarding the nature of the dark sector and are a fundamental tool in order to test the theoretical consistency and the phenomenological consequences of modified or quantum gravity theories \cite{PerezBergliaffa:2011gj, Olmo2015, Bonanno2000, Bonanno2006, Ishibashi2021, Modesto_2006, Bambi2013, Akil2023}. Primordial black holes have long been thought to be responsible for at least part of the Dark Matter we observe in our universe \cite{Belotsky2014, Yang2020pbhasDM}. Recently, black holes have also been connected with dark energy, with models of cosmologically coupled black holes being used to explain the acceleration of the expansion of the universe \cite{Croker_2020, Farrah_2023, Cadoni_2023, DESICCBHs2025}.

Until very recently, black holes themselves were considered to be theoretical structures supported only by indirect evidence \cite{Mitra2000, Schild_2006, Berti_2015}. However, the observation of gravitational waves that can be perfectly explained by the coalescence of black hole binaries \cite{LIGOScientific:2020ibl} and the observation of black hole shadows by the Event Horizon Telescope collaboration \cite{Akiyama_2019, EventHorizonTelescope:2022wkp} has granted this class of Einstein's equations solutions the status of physically real. That being stated, there is still a strong debate surrounding the physics of event horizons and the existence of singularities which is directly connected with the possible existence of wormholes \cite{Berti_2015,Ashtekar2022, Simpson_2019}. The ultimate judge of those debates is observation and experiment, so also in the context of non-singular black hole solutions or horizon-less compact objects that mimics them, the literature is mostly focused on deriving predictions in specific contexts and theoretical models \cite{Olmo2015, Abdujabbarov2016, Annulli_2022, Carballo-Rubio_2022}. In that direction, there has been a growth in interest on developing more precise descriptions of black holes and their environment, ranging from a better understanding of the process of accretion by primordial black holes \cite{Loeb_2024, Facchinetti2023, Jangra_2025}, passing through the impact of supermassive black holes in dark matter halos and filaments of galaxies \cite{Cardoso2023, vgn2025, Jung2025}, reaching the possibility that cosmologically coupled black holes could simultaneously explain the recent detection of active galactic nuclei by the James Webb telescope (commonly referred to as ``little red dots'' \cite{Matthee_2024}) and impact the accelerated expansion rate of the universe \cite{Croker_2020, Farrah_2023}. It is possible that both black holes and wormholes inhabit our universe and we are simply not able to differentiate between their physical effects at this point. If that is the case and both those structures are impacting the astrophysics and cosmology of our universe, it is necessary to derive the exact consequences of these geometric structures in evermore realistic physical situations.

Considering the current challenges facing the standard cosmological model, a better understanding of the interaction between compact objects and their astrophysical environments is of fundamental importance to improve our understanding of cosmic evolution. Moreover, a more complete and exact understanding of how the phenomenology of compact objects is impacted by different environments can put stronger constraints on their fundamental nature. The Dark Matter ``mystery'' \cite{cirelli2024darkmatter} (namely the problem of defining its nature beyond the dust fluid approximation) can also benefit from these improved models by crosschecking cosmological and astrophysical phenomenology from different models and theories. In \cite{Cardoso2023, vgn2025} it was demonstrated that different Dark Matter density profiles will directly influence the phenomenological observables related to supermassive black holes at their center. As it was noted in \cite{vgn2025}, their work can also present an opportunity to test different Dark Matter models through their effects on black hole shadows or gravitational lensing. Some works have investigated whether the most commonly used dark matter profiles, namely the Navarro, Frenk, and White (NFW) model \cite{NFW96} and the Universal Rotation Curve (URC) profile \cite{URC2012}, exhibit the intrinsic characteristics necessary to support traversable wormholes \cite{Hassan2024}. Other works have focused on specific density distributions and modified theories of gravity. For instance, the Dekel-Zhao (DZ) dark matter profile \cite{Zhao1996,Zhao1997,Freundlich2020} has been effectively employed to derive wormhole solutions \cite{Khatri2025}.



The existence of wormholes will probably depend on the fundamental nature of the gravitational field \cite{Maldacena2013, Capozziello:2022zoz, Cruz_2024}, since no matter what path it takes to its formation it will probably involve high energy processes at scales beyond the scope of General Relativity. Developing a Quantum Theory of Gravity remains one of the central problems of modern physics \cite{Bambi2024}, although some frameworks like Loop Quantum Gravity (LQG) have gained increased popularity and have been applied to multiple scenarios where singularities are avoided due to quantum effects at high energies and small distances \cite{Modesto_2006, Bojowald2020}. In that framework one can construct wormhole solutions and those solutions have been studied in a variety of contexts with interesting results \cite{Cruz_2024}. Although LQG and String Theory remain the most popular approaches to the problem of quantizing the gravitational field, both of them suffer from similar limitations, such as the lack of decisive experimental predictions and the difficulty in describing how spacetime emerges with matter from a more fundamental unified structure \cite{Rovelli_2020, Huggett_Vistarini_2015}. In recent years, a number of authors have proposed different approaches to what is sometimes called ``Quantum General Relativity" (QGR) or ``Quantum Einstein Gravity " \cite{Donoghue2012, Reuter2002}. Some authors have argued that you can apply direct quantum field theory methods to the Einstein-Hilbert (EH) action and, assuming that there is some ultraviolet cutoff scale that is well beyond the scale of any experiment, obtain perturbative corrections that introduce scale dependence in the fundamental coupling constants (the Newton constant $G$ and the cosmological constant $\Lambda$) \cite{Donoghue2012}. Others have argued in favor of ``super-renormalizable" versions of quantum gravity, introducing higher derivative terms in the action in a way that might be free of instabilities thought to plague this kind of quantum theory \cite{Modesto2012, Giacchini2020, Asorey:2024mkb}. Another approach that has gained popularity in recent years is also based on quantum field theory methods but tries to derive the exact form of the running from numerical calculations of the renormalization group behavior of QGR \cite{Reuter2002, PhysRevD.57.971, Eichhorn2024}. This is known as the Asymptotically Safe Gravity (ASG) approach and is based on the conjecture that Einstein's theory is renormalizable at some scale and we just need to find the correct fixed point $\beta$ for the renormalization group flow of the coupling constants, as is done in other quantum field theories. This method can be used to derive a precise dependence of $G=G(k)$ with the mass scale $k$ of the infrared cutoff and with that exact function one can try to ``improve" the equations of general relativity to derive possible effects of this scale dependence on the physics of the gravitational field \cite{Bonanno2000, Bonanno2006, PhysRevLett.123.101301, Ishibashi2021}. This improvement can be done in different ways: i) inserting $G(k)$ (and $\Lambda(k)$ when needed) directly into the action and obtaining modified field equations; ii) inserting the running(s) directly into Einstein's equation and deriving modified solutions with a scale dependent metric and energy-momentum tensor\footnote{A complete treatment in ASG would require to consider the running of the couplings of the matter degrees of freedom also. In this work we will work only with the dust fluid approximation for DM and will not consider possible corrections from ASG in the nature of DM itself, although this is an active and interesting subject being explored in the literature \cite{Brito2024}.} or iii) inserting $G(k)$ and $\Lambda(k)$ into known solutions of Einstein's equations. Besides that, one also has to make a ``cutoff identification'' defining how the momentum dependence of the Newton constant will translate into spacetime. There is no generally agreed upon definition for these last identification, however Bonanno and Reuter \cite{Bonanno2006} have argued that, for the case of the Schwarszchild black hole at least, it is enough to take $k(r) = \xi/r$ to obtain quantum corrections to the usual solution that are consistent with more general choices (in leading order). A more complete discussion on the different improvement and cutoff choices can be found in \cite{Ishibashi2021}, where the analysis of the charged black hole led the authors to argue that a general relation of the type $k \sim \xi/r^p$ is consistent with a more general cutoff identification involving the Kretschmann scalar. As was mentioned, ASG has also been used to describe quantum improved space-times of wormholes \cite{Alencar2021, Nilton2021, Nilton2022} showing the potential of this approach to obtain new phenomenological predictions related to quantum gravity and exotic astrophysical objects. In these works the method of improvement was through the modification of the EH action and the cutoff identification was left generic for part of the derivation, using a $G(\chi)$ function where $\chi$ was assumed to be any curvature invariant (such as the Ricci, Riemman or Kretschmann scalars). In the case of the Ellis-Bronikov wormhole studied in \cite{Alencar2021} the identification ultimately led to a similar result as in \cite{Ishibashi2021} but with $p=2$. In the cases analyzed in \cite{Nilton2021, Nilton2022}, however, the cutoff function $f(\chi)$ (that is related to $k$ through $\omega k^2$ $\rightarrow$ $\xi f(\chi)$) is proportional to the Ricci scalar which in turn is proportional to $r^{-2}$ leading to a cutoff identification of the type $k \sim 1/r$.

In this work we will employ the quantum improvement directly into Einstein's equations for a static and spherically symmetric spacetime sourced by a DM density profile of the Dekel-Zhao type \cite{Zhao1996,Zhao1997,Freundlich2020, Khatri2025}. We will make the simple cutoff identification $k(r) = \xi/r$ which is consistent with the IR limit and asymptotical flatness, and is also justified by the cited works involving improved black hole solutions \cite{Bonanno2000, Bonanno2006, PhysRevLett.123.101301, Ishibashi2021} and wormholes \cite{Nilton2021, Nilton2022}. We are mostly interested in investigating the necessity of exotic matter to guarantee the structures stability and will analyze this through the energy conditions and the solutions of the Tolman–Oppenheimer–Volkoff (TOV) equation. We will see that the running of the gravitational constant introduces a fundamentally quantum effect, providing a type of repulsive force that ensures the stability of the throat but still requires that the matter density be exotic.




The organization of this paper is outlined as follows. In Section \ref{secASG}, we discuss the elements of our model including the justification for our position dependent Newton constant, how it is introduced directly into Einstein's field equations, as well as the basic properties of generic traversable Lorentzian wormholes and of the Dekel-Zhao dark matter profile. Section \ref{secGF} presents the geometric features of an ASG wormhole sourced by dark matter, including shape and redshift functions, geometric constraints, curvature aspects, and embedding diagrams. Sections \ref{secPF} and \ref{secPheF} delve into the physical and phenomenological features of this construction, including energy and equilibrium conditions, as well as the shadows of an ASG wormhole. Finally, Section \ref{seccon} presents a discussion of our results.

\section{ASG and traversable wormholes sourced by dark matter}\label{secASG}

In this section we introduce our improved wormhole solution within the Dekel-Zhao DM halo density profile. We adopt here an effective formulation of ASG by introducing the scale-dependent gravitational coupling directly into the field equations, rather than in the action or in the metric. This implementation preserves the theoretical structure of General Relativity while consistently incorporating the infrared running of the coupling, at least to leading order. This choice is made both based on the previously mentioned results \cite{Bonanno2006, Ishibashi2021, Nilton2021, Nilton2022} and to allow for analytical calculations of the relevant geometrical quantities. We shall see that it leads to an interesting scenario with some of the expected features of quantum gravity and is appropriate for the relevant astrophysical scales of compact objects and DM halos. We introduce the scale dependent Newton constant and the modified equations of motion in Subsection \ref{ASG} and briefly discuss the general properties of TWs and the Dekel-Zhao DM density profile in the following subsections. With those elements, we will be prepared to study the geometrical properties of the quantum improved wormhole in Section \ref{secGF}.

\subsection{Asymptotic safety in Quantum Gravity}\label{ASG}

The framework of ASG is founded on the hypothesis that the Einstein--Hilbert theory admits a non-Gaussian fixed point (NGFP) in the renormalization group (RG) flow of its coupling constants~\cite{PhysRevD.57.971, Eichhorn2024}. As a result, the gravitational couplings become scale-dependent quantities, such that \( G \to G(k) \) and \( \Lambda \to \Lambda(k) \), instead of universal constants. Numerical analysis of the renormalization group equation in the case of pure gravity with the Einstein-Hilbert truncation has led several authors to the functional form \cite{Bonanno2000}
\begin{equation}
    G(k) = \frac{G_0}{1+\omega G_0 k^2},
\end{equation}
where $k$ represents the momentum scale where quantum fluctuations affect the value of the gravitational coupling, $\omega= \frac{4}{\pi}(1-\frac{\pi^2}{144})$ and $G_0$ is the standard value of the Newton constant. The next step is to identify the cutoff scale $k$ with some physical length that is relevant to the problem we are interested in. In the infrared regime relevant for galactic scales, the identification \(k \propto r^{-1}\) leads to
\begin{equation}
G(r) = \frac{G_0\, r^2}{r^2 + \xi^2},
\label{eq:G_r}
\end{equation}
where \(\xi\) is the characteristic length scale associated with the ASG corrections and  $\omega$ is absorbed into $\xi$. As a first approximation in the infrared regime, this modification provides a starting point to investigate how ASG corrections alter the spacetime geometry and whether they can mitigate the need for exotic matter to sustain a traversable wormhole within a dark matter halo. Substituting Eq.~(\ref{eq:G_r}) into Einstein's equations gives the \textit{ASG-modified Einstein equations} used in this work,
\begin{equation}
G_{\mu\nu} = 8\pi\, G(r)\, T_{\mu\nu}^{(\mathrm{DM})}.
\label{eq:Einstein_ASG}
\end{equation}
When the classical source is a galactic dark matter halo described by the Dekel-Zhao (DZ) density profile, the interplay between the ASG corrections and the matter density \(\rho(r)\) defines the effective curvature. It is worth noticing that the standard Einstein equations are recovered in the classical limit \(\xi \rightarrow 0\), whereas in the quantum regime the geometry becomes \textit{self-regulated}~\cite{PhysRevLett.123.101301}. In the present context of IR regime, such behavior allows for regular wormhole solutions that satisfy both the flare-out condition and asymptotic flatness, as discussed in the following sections.

\subsection{Traversable Lorentzian wormholes}\label{TWSlorentz}

The geometric framework for a traversable wormhole is established using the Morris-Thorne metric, which describes a static, spherically symmetric spacetime \cite{Morris1988},
\begin{equation}\label{metric}
ds^2 = -e^{2\Phi(r)}\,dt^2 + \frac{dr^2}{1 - \frac{S(r)}{r}} + r^2\bigl(d\theta^2 + \sin^2\theta\,d\phi^2\bigr)\ ,
\end{equation}
where $t$ is the temporal coordinate, $r$ the radial coordinate, and $\theta$ and $\phi$ the angular coordinates. This metric is characterized by two principal functions: the redshift function, $\Phi(r)$, which governs gravitational time dilation, and the shape function, $S(r)$, which determines the spatial geometry of the wormhole. For the wormhole to be traversable, the redshift function must remain finite everywhere. This requirement is critical because it ensures the absence of event horizons—surfaces beyond which no return is possible, thus rendering two-way travel impossible. The shape function corresponds to the wormhole's spatial geometry and must satisfy several conditions to ensure both stability and traversability. It molds the spatial curvature and must obey a strict set of conditions to ensure the structure is physically coherent and stable. The geometric constraints imposed by the shape function determine the traversability of the wormhole. The first and most basic condition defines the wormhole's throat at $r=r_0$, where $r_0$ is the throat's radius. Mathematically, this constraint is expressed as $S(r_0) = r_0$. This equality establishes the minimal circumference of the structure, serving as the boundary condition for the entire geometry. The most critical condition is the ``flare-out''. Geometrically, it is expressed as $S(r)-S'(r)r>0$, and must be satisfied for any $r$ value. This inequality ensures that the throat's surface curves outward in both directions. Without this property, the wormhole would not effectively open. Within the throat, the ``flare-out'' condition can be expressed as $S'(r_0)\leq 1$. This condition has a direct physical interpretation related to tidal forces. The derivative $S'(r)$ describes the rate at which the spacetime geometry changes. If the geometry were to curve too sharply at the throat (i.e., if $S'(r_0) > 1$), the resulting gravitational tidal forces would be extraordinarily high. For the entire region outside the throat ($r>r_0$), the relation $S(r)/r < 1$ must hold. This ensures that the radial component of the metric remains positive and finite. Finally, the asymptotic flatness condition requires that $S(r)/r \rightarrow 0$ as $r \rightarrow \infty$. This condition ensures that at large distances from the wormhole, the spacetime is flat. However, in some cases, this condition can be relaxed \cite{Sultan2025,Mustafa2023}.   

\subsection{The Dekel-Zhao dark matter profile}

The general form of Dekel-Zhao dark matter density profile is expressed as \cite{Zhao1996,Zhao1997},
\begin{equation}\label{dzrho}
\rho(r)=\frac{\rho_0}{\left(\frac{r}{r_c}\right)^a \left[1 + \left(\frac{r}{r_c}\right)^{1/b}\right]^{b(\gamma - a)}},
\end{equation}
where $r$ is the radial distance, $r_c$ is the characteristic radius of the density profile, and $\rho_0$ is a reference/characteristic density. The last two parameters are associated with the quantity of dark matter. Thus, $r/r_c$ represents the normalized radial distance. The parameters $a,b$ and $\gamma$ govern the inner slope, transition sharpness, and outer slope of the density profile, respectively \cite{Khatri2025}. By adjusting its defining parameters, the profile can accurately model a diverse range of empirically observed dark matter distributions. This adaptability elucidates the fundamental characteristics and dynamics of dark matter in the universe. Our mathematical representation of the DZ profile is formulated by setting $b=1$ and $\gamma = 3$.
\begin{equation}\label{dzrho2}
\rho(r)=\frac{\rho_0}{\left(\frac{r}{r_c}\right)^a \left[1 + \left(\frac{r}{r_c}\right)\right]^{(3 - a)}}.
\end{equation}

\begin{figure}[!htp]
\centering
\includegraphics[width=1\textwidth]{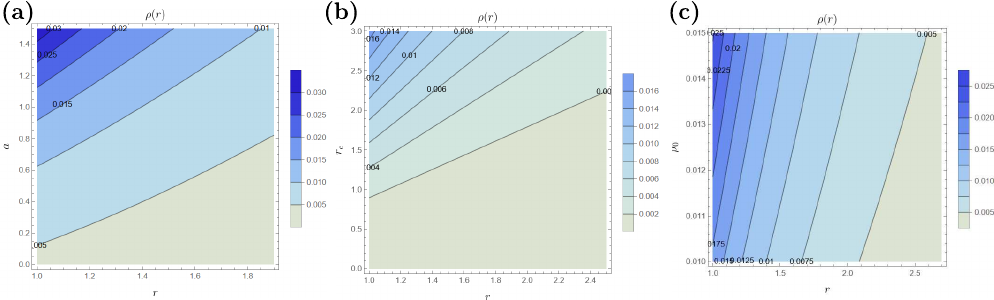}
\caption{The Dekel-Zhao's dark matter density profile, $\rho$, for different keys parameters as a function of $r$, with: (a) $\rho_0 = 0.01$ and $r_c =3.0$. (b) $\rho_0 = 0.01$ and $a =1.0$. (c) $a = 1.0$ and $r_c =3$. The $b = 1$ and $\gamma = 3$ are fixed.  }\label{figrho}
\end{figure}

The behavior of the DZ dark matter density profile, \eqref{dzrho2}, as a function of radial distance $r$ and key parameters ($a,r_c,\rho_0$), is illustrated in Fig. \ref{figrho}[a-c]. The profile is maintained within the NFW-like regime by using fixed values of $b=1$ and $\gamma = 3$. These plots are essential for understanding how the structural properties of the dark matter halo influence the local matter density, which is a  critical source term for the spacetime geometry.

The plots demonstrate a high sensitivity of the density profile to its defining parameters. Panel \ref{figrho}(a) explores the influence of the inner slope parameter, $a$, and its relationship to the quantity of dark matter in the halo. A higher $a$ corresponds to a more centrally concentrated dark matter distribution. In panel \ref{figrho}(b) we investigate the role of the characteristic radius, $r_c$. The results indicate that an increase in $r_c$ leads to a higher density at a given inner radial distance. This suggests that dark matter halos with small $r_c$ are compact. For a wormhole, $r_c$ must exceed the throat radius for physical consistency. Finally, panel \ref{figrho}(c) demonstrates the linear relationship between $\rho(r)$ and $\rho_0$, as expected from \eqref{dzrho2}. The three parameters are fundamental to our analysis, as the high sensitivity of the \eqref{dzrho2} to its defining characteristics can significantly impact geometry and, consequently, the viability of the wormhole.

\section{Geometric features}\label{secGF}

In this section, we examine the geometric structure of the traversable wormhole supported by the Dekel-Zhao dark matter halo within the framework of ASG. We derive and analyze the modified shape and redshift functions resulting from the running gravitational coupling $G(r)$, and investigate the geometric constraints that ensure a physically viable configuration. The analysis further explores the curvature properties, highlighting how the ASG parameter $\xi$ and the dark matter distribution parameters $(a,\,\rho_{0},\,r_{c})$ shape the local and global features of spacetime. Finally, embedding diagrams are constructed to visualize the wormhole geometry, offering an intuitive picture of how quantum-gravity corrections and dark matter jointly influence the curvature and traversability of the structure.

\subsection{Shape and redshift functions}

We will now investigate the wormhole solutions in the context of our ASG improved Einstein's equations. We will particularly focus on the energy density given in Eq. \eqref{dzrho2} by the static and spherically symmetric Morris-Thorne wormhole metric as presented in Eq. (\ref{metric}). Thus, the modified Einstein equations take on their simplest form:
\begin{align}
G_{\ t}^{t} = & \frac{S'}{r^{2}}=8\pi G(r)\rho(r),\label{eq:g00}\\
G_{\ r}^{r} = & -\frac{S}{r^{3}}+ 2\frac{\left(r-S\right)\Phi'}{r^{2}}= 8\pi G(r) P_r(r),\label{eq:grr}\\
G_{\theta}^{\theta} = & G_{\phi}^{\phi} = \left(1-\frac{S}{r}\right)\left[\Phi''+(\Phi')^{2}+\frac{\left(S-rS'\right)}{2r(r-S)}\Phi'+\frac{\left(S-rS'\right)}{2r^2(r-S)}+\frac{\Phi'}{r}\right]= 8\pi G(r)P_{t}(r)\label{eq:gthetatheta},
\end{align}
where $P_r$ and $P_t$ are radial and tangential pressures, respectively.

Integrating Eq. (\ref{eq:g00}), we have that the shape function is given by
\begin{equation}
S(r)=c_1+8\pi\int G(r)\rho(r) r^2 dr,
\end{equation}
where $c_1$ is the integration constant, which can be found through the boundary condition $S(r_0)=r_0$ at the throat. Thus, the shape function becomes, on considering Eq. (\ref{eq:G_r}):
\begin{equation}\label{defshape}
   S(r)=r_0-s(r_0)+s(r),
\end{equation}
where
\begin{eqnarray}\label{shapefunction}
s(r)&=& \frac{4\pi
\rho_0 \, r^3 \left(\frac{r}{r_c}\right)^{-a}
}{
 (a - 3)(r + r_c)^3
}
\Bigg\{
r_c^3 \left(\frac{r + r_c}{r_c}\right)^a
{}_2F_1\left[1, 3 - a; 4 - a; \frac{r\left(1 + i \frac{r_c}{\xi}\right)}{r + r_c} \right]
\nonumber \\
&& \quad +
r_c^3 \left(\frac{r + r_c}{r_c}\right)^a
{}_2F_1\left[1, 3 - a; 4 - a; \frac{r\left(1 - i \frac{r_c}{\xi}\right)}{r + r_c} \right]
\quad -
2 (r + r_c)^3
{}_2F_1\left(3 - a, 3 - a; 4 - a; -\frac{r}{r_c} \right)
\Bigg\}.
\end{eqnarray}
Here, we have fixed the parameters $b=1$ and $\gamma=3$. Although some Gauss hypergeometric functions involve complex arguments, the corresponding terms are complex conjugates, resulting in a real-valued sum. For instance, considering $a=1$ (NFW dark matter profile) we have
\begin{eqnarray}
s(r) &=& \frac{4\pi \rho_0 \, r_c^3 }{ (r_c^2 + \xi^2)^2} \, \Bigg[ 
\frac{2 r_c^3 (r_c^2 + \xi^2)}{r + r_c}
- 4 r_c \, \xi^3 \arctan\left( \frac{r}{\xi} \right) \nonumber \\
&& \quad + 2 (r_c^4 + 3 r_c^2 \xi^2) 
\log\left( \frac{r + r_c}{r_0 + r_c} \right)
+ (-r_c^2 \xi^2 + \xi^4) 
\log\left( \frac{r^2 + \xi^2}{r_0^2 + \xi^2} \right) 
\Bigg].
\end{eqnarray}

It is important to stress that the calculation of the shape function through the Einstein equations is strongly affected by the scale-dependent gravitational coupling. The variation of $G(r)$ motivated by ASG introduces significant corrections to the shape function when compared with the constant coupling of General Relativity. As a consequence, the geometric and physical properties of the wormhole are substantially modified. This highlights the nontrivial role played by quantum corrections in determining the effective wormhole configuration.

\begin{figure}[!htp]
\centering
\includegraphics[width=1\textwidth]{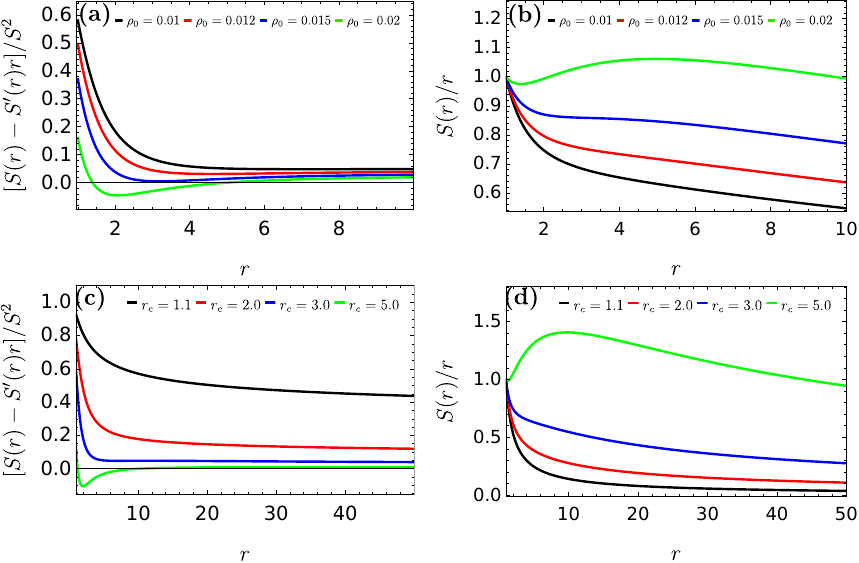}
\caption{The flare-out, $[S(r)-S'(r)r]/S^2$, and asymptotic conditions with variations of $\rho_0$ and $r_c$, for $a=1.0$, $\xi = 0.1$ and $r_0 = 1.0$ fixed. (a)-(b) $\rho_0=0.01,\,0.012,\,0.015, \,0.02$ ($r_c=3.0$) and (c)-(d)$r_c=1.1,\,2.0,\,3.0,\,5.0$ ($\rho_0=0.01$) .}\label{fig1}
\end{figure}

\begin{figure}[!htp]
\centering
\includegraphics[width=1\textwidth]{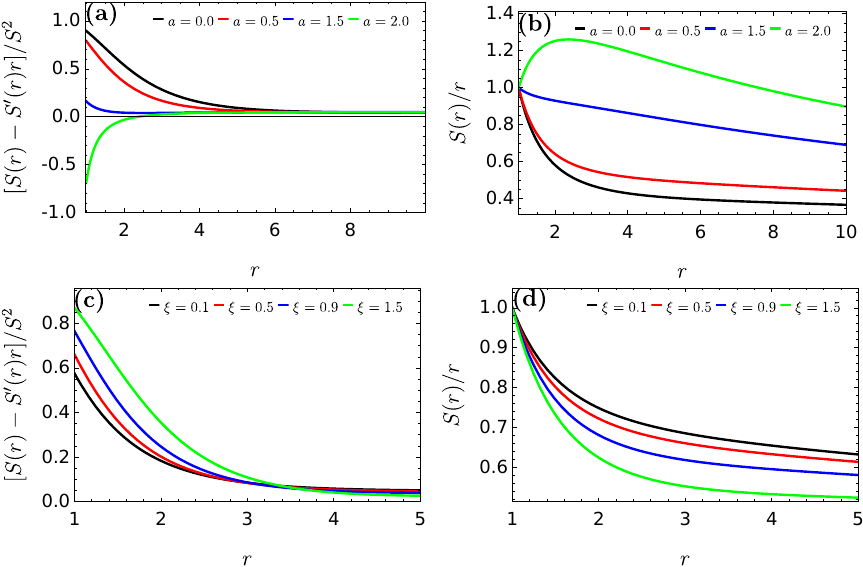}
\caption{The flare-out, $[S(r)-S'(r)r]/S^2$, and asymptotic conditions with variations of $a$ and $\xi$, for $\rho_0=0.01$, $r_c = 3$ and $r_0 = 1.0$ fixed. (a)-(b) $a=0.0,\,0.5, \,1.5, \,2.0$ ($\xi=0.1$) and (c)-(d) $\xi=0.1,\,0.5,\,0.9,\,1.5$ ($a=1.0$) .}\label{fig2}
\end{figure}

Figures \ref{fig1} and \ref{fig2} illustrate the geometric conditions required for the existence of the wormhole. The results show that these conditions are highly sensitive to the parameters of the dark matter distribution. In particular, a larger characteristic density $\rho_0$ tends to obstruct the formation of the wormhole, since the increased dark matter concentration strongly modifies the shape function and prevents the flare--out condition from being satisfied. Similarly, the scale radius $r_c$ plays a crucial role: an extended dark matter halo contributes to suppressing the wormhole geometry. Furthermore, deviations from the standard NFW profile ($a=1$) also disfavor the wormhole formation, indicating that the existence of a traversable configuration is tightly constrained to the case where the dark matter halo follows closely the NFW distribution. In contrast, variations in the ASG parameter $\xi$ do not prevent the formation of the wormhole, although they significantly affect the local geometry, especially by modifying the curvature behavior in the vicinity of the throat. These findings emphasize the fine tuning between the dark matter parameters and the geometric conditions required to sustain a wormhole throat.

Regarding the redshift function, we will adopt
\begin{equation}\label{redshift}
\Phi(r) = 2\ln\left(\frac{r}{r+r_0}\right).
\end{equation}
The choice for this function is physically well motivated, since it remains finite at the throat, where its constant value can be absorbed into a redefinition of the time coordinate, and it vanishes in the asymptotic limit $r \to \infty$, ensuring consistency with a flat background. Moreover, this form guarantees the absence of horizons, since $g_{tt}=e^{2\Phi(r)}$ never vanishes in the domain $r_0\leq r<\infty$, and provides smooth derivatives that avoid divergences in the pressure components. An additional advantage is that the logarithmic dependence helps to alleviate the violation of the energy conditions by softening the terms involving $\Phi'(r)$ in the Einstein equations. Finally, its analytical simplicity allows explicit evaluation of curvature scalars, embedding diagrams, and stability conditions, while offering a more physically rich structure than the trivial choice $\Phi(r)=0$ often adopted in the literature.

\subsection{Curvatures}

\begin{figure}[!htp]
\centering
\includegraphics[width=1\textwidth]{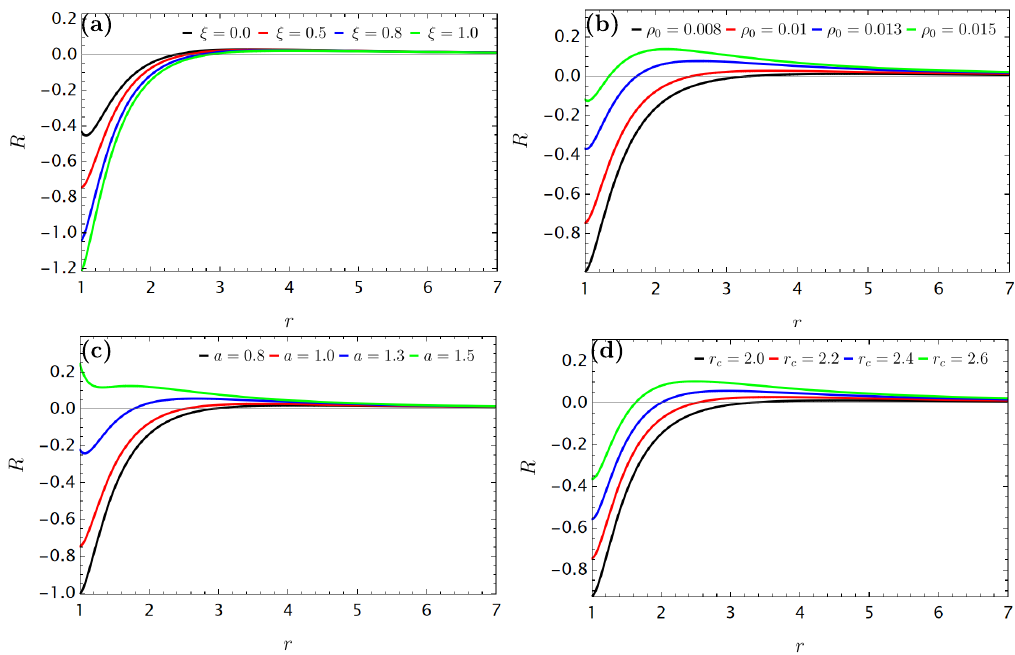}
\caption{Ricci's scalar as a function of the radial coordinate, produced for Dekel-Zhao's dark matter density profile: (a) $\xi = 0, 0.5, 0.8, 1.0$($a=1.0, r_c= 2.2, \rho_0 = 0.01)$. (b) $\rho_0 = 0.008, 0.01, 0.013, 0.015$ ($a = 1.0, r_c = 2.2, \xi = 0.5$). (c) $a = 0.8, 1.0, 1.3, 1.5$ ($\rho_0 = 0.01, r_c = 2.2, \xi = 0.5$). (d) $r_c = 2.0, 2.2, 2.4, 2.6$ ($a=1.0, \rho_0 = 0.01, \xi = 0.5$). In all panels $r_0 = 1.0$.}\label{fig9}
\end{figure}

The analysis of the Ricci scalar curvature, depicted in Fig. \ref{fig9}, unveils key aspects of the geometric structure of the wormhole spacetime under the Dekel-Zhao dark matter profile with ASG corrections. Panel~(a) highlights the role of the ASG parameter \(\xi\) in shaping curvature behavior. For smaller values of \(\xi\), the Ricci scalar displays a milder negative peak near the throat region (\(r \approx r_0 = 1.0\)), indicating weaker curvature concentration. As \(\xi\) increases, this peak sharpens considerably, signaling an enhancement of quantum gravity effects. This transition reflects the smoothing influence of ASG corrections, which effectively regulate the gravitational coupling in the high-curvature region close to the throat. Such a behavior motivates the identification of the ASG energy scale with the throat curvature, a point we explore in greater detail in the shadow calculation. Since the throat constitutes the region of maximal curvature and represents the most distinctive geometric feature of the wormhole, it naturally provides a physically meaningful scale for quantum gravity corrections.

 Panels~(b) and~(d) illustrate the dependence of the Ricci scalar curvature on the central dark matter density \(\rho_0\) and the characteristic scale \(r_c\), respectively. Interestingly, at the throat (\(r \approx r_0\)), larger values of \(\rho_0\) or \(r_c\) result in a less negative Ricci scalar, indicating a local smoothing of the curvature in this region. However, in the vicinity of the throat, there emerges a local maximum of curvature that is positive, whose magnitude increases with \(\rho_0\) and \(r_c\). This behavior reflects the interplay between the throat geometry and the dark matter distribution: higher \(\rho_0\) and larger \(r_c\) enhance the gravitational influence of the Dekel-Zhao dark matter profile, producing a stronger curvature peak near—but not exactly at—the throat.

The influence of the inner slope parameter \(a\) is illustrated in panel~(c). For the standard NFW profile (\(a = 1.0\)), the Ricci scalar exhibits a characteristic curvature profile with moderately negative values. Deviations from this canonical case lead to noticeable geometric variations: smaller \(a\) values (\(a = 0.8\)) produce more negative curvature, indicating a sharper geometric transition near the throat, whereas larger \(a\) values (\(a = 1.3, 1.5\)) yield less negative curvature, corresponding to smoother and more diffuse curvature distributions. This pronounced dependence on the inner slope parameter demonstrates how the detailed structure of the dark matter density profile directly governs the wormhole’s curvature properties.

\subsection{Embedding diagrams}

\begin{figure}[!htp]
\centering
\includegraphics[width=1\textwidth]{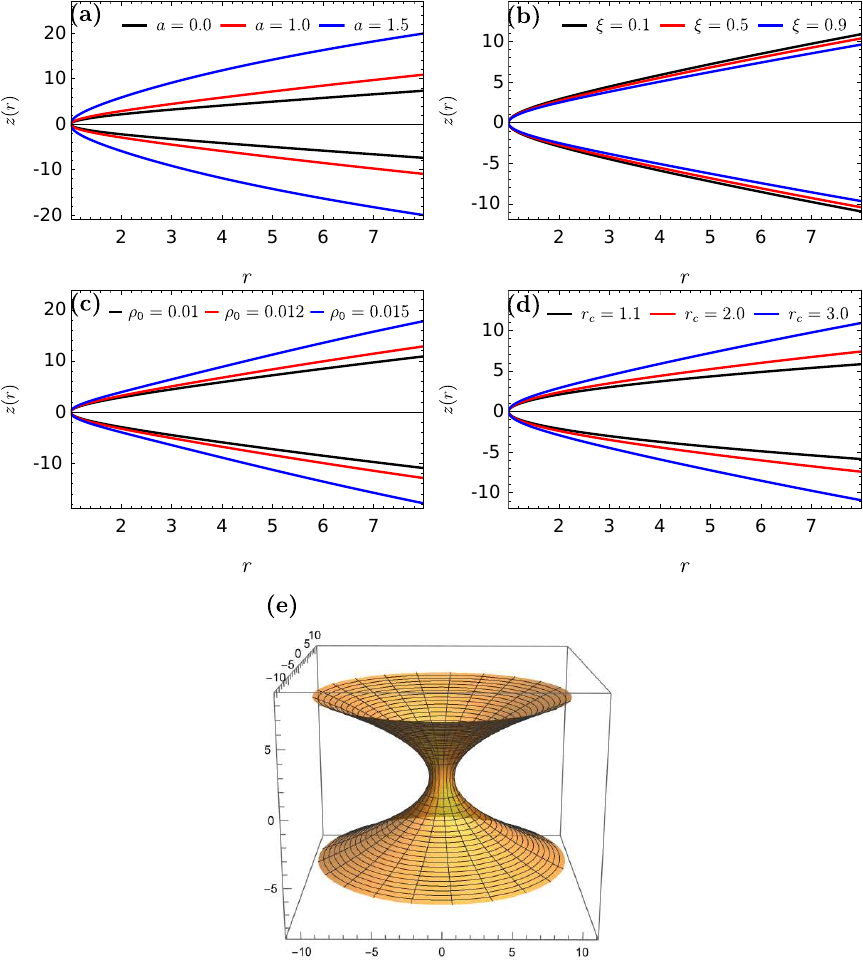}
\caption{Embedding diagrams of a wormhole produced for Dekel-Zhao's dark matter density profile. (a) $a=0.0,\,1.0,\,1.5$ ($\xi=0.1,\ \rho_0=0.01,\ r_c=3.0$). (b) $\xi=0.1,\,0.5,\,0.9$ ($a=1.0,\ \rho_0=0.01,\ r_c=3.0$). (c) $\rho_0=0.01,\,0.012,\,0.015$ ($a=1.0,\ \xi=0.1,\ r_c=3.0$). (d) $r_c=1.1,\,2.0,\,3.0$ ($a=1.0,\ \xi=0.1,\ \rho_0=0.01$). (e) The 3D embedding for $a=1.0,\ \xi=0.1,\ \rho_0=0.01,\ r_c=1.1$. In all panels $r_0=1.0$.}\label{fig3}
\end{figure}

The spatial geometry of a static and spherically symmetric wormhole is typically represented using an embedding diagram. This technique effectively illustrates how the spatial structure of the wormhole throat is embedded within a higher-dimensional space, providing essential visual clarity. The analysis begins by simplifying the four-dimensional spacetime to a more manageable two-dimensional surface, resulting in minimal loss of geometric information. By considering a static configuration with time held constant and focusing on an equatorial slice ($\theta = \pi/2$), the line element for this slice is given by
\begin{equation}\label{linelesta}
    ds^2 = \frac{dr^2}{1-\frac{S(r)}{r}}+r^2d\phi^2.
\end{equation}
The objective is to embed this two-dimensional curved manifold into a flat, three-dimensional Euclidean space. In cylindrical coordinates ($r,\phi,z$), the metric of this embedding space is expressed as
\begin{equation}
    ds^2=dz^2+dr^2+r^2d\phi^2.
\end{equation}
The embedded surface is designed to exhibit axial symmetry, allowing it to be described by a single function $z=z(r)$. Thus, when restating the line element in cylindrical coordinates, we have
\begin{equation}\label{linelstac}
    ds^2 = \left[1+\left(\frac{dz}{dr}\right)^2\right]dr^2+r^2d\phi^2.
\end{equation}
By ensuring that the geometry of the equatorial slice is identical to that of the embedded surface, we can equate the two expressions for $ds^2$, \eqref{linelesta} and \eqref{linelstac}. This identification results in the fundamental differential equation that governs the shape of the embedded surface.
\begin{equation}
    \frac{dz}{dr} = \pm \sqrt{\frac{S(r)}{r-S(r)}}.
\end{equation}
This equation clearly illustrates how the spatial geometry of the wormhole is influenced by the chosen shape function, $S(r)$, as depicted in Figure \ref{fig3}[a-d]. The resulting 3D visualization shows two distinct, asymptotically flat regions of space connected by a throat, which corresponds to the location of the minimum radius, Figure \ref{fig3}(e).

More than just an illustration, the diagram offers valuable insights into the structure of the wormhole. The resulting visual displays a smooth transition from the throat to the asymptotic regions, indicating that the geometry does not exhibit sharp discontinuities or singularities. This well-defined curvature profile supports the physical viability of the solution, reinforcing that the shape function has been chosen in a manner that allows for a smooth and potentially traversable wormhole structure. Consequently, the embedding diagram acts as a powerful diagnostic tool, visually confirming the geometric integrity of the spacetime.

Figure \ref{fig3} presents the embedding diagrams for different choices of the dark matter parameters. When the concentration or extent of dark matter is small -- see panels (c) and (d) -- the throat appears sharper and the spatial curvature at $r=r_0$ is more pronounced, giving rise to a ``pointed'' geometry. In these cases, however, the wormhole quickly approaches the asymptotically flat region, since the influence of dark matter on the geometry is weak away from the throat. On the other hand, when the dark matter concentration or extent is increased, the throat becomes smoother and less curved locally, but the spatial curvature extends further from the throat, and the wormhole requires a larger radial distance to recover flatness. This behavior indicates that dark matter acts as an extended gravitational source: it reduces the sharpness of the curvature at the throat while simultaneously spreading the deformation of the geometry over larger scales.

The ASG parameter also has a distinctive effect on the embedding geometry. As shown in panel (b) of the same figure, increasing $\xi$ makes the spatial curvature of the throat at $r=r_0$ sharper. This occurs because a larger $\xi$ suppresses the effective gravitational coupling $G(r)$ in the vicinity of the throat, reducing the gravitational contribution of the matter sources and forcing the curvature to concentrate locally in order to sustain the wormhole opening. On the other hand, since $G(r) \to 1$ at large distances, the recovery of the classical coupling occurs more rapidly, and the wormhole flattens to its asymptotically flat geometry at shorter radial scales. Therefore, the ASG parameter acts as a regulator that both accentuates the local curvature at the throat and simultaneously favors a faster transition to flatness away from it.

\section{Physical features}\label{secPF}

Having established the geometric foundations of our wormhole model, we now turn to an examination of its physical characteristics. This section investigates the fundamental physical viability of the ASG-corrected wormhole sourced by Dekel-Zhao dark matter, focusing specifically on the energy conditions that govern matter content and the equilibrium conditions that determine stability. These analyses are crucial for assessing whether such theoretical constructs could represent physically plausible objects within the framework of ASG.

\subsection{Energy conditions}

Energy conditions are a fundamental topic in General Relativity, essential for ensuring that the theory aligns with the physically coherent behavior of matter and energy. These conditions impose restrictions on the energy-momentum tensor, $T_{\mu\nu}$, to ensure that energy is always non-negative and that causality is preserved in the physical universe \cite{kontou2020energy}. Rigorously, the energy conditions are not provable theorems; they are restrictions on the energy-momentum tensor that are assumed to be valid for all physically plausible forms of matter. The principal conditions include the Weak Energy Condition (WEC), the Strong Energy Condition (SEC), the Dominant Energy Condition (DEC) and the Null Energy Condition (NEC). The first condition, WEC, states that the energy density measured by any observer along a timelike worldline must be non-negative,
\begin{equation}
    T_{\mu\nu}t^\mu t^\nu \ge 0,
\end{equation}
where $t^\mu$ is a timelike vector. The SEC requires that the effective energy density, as measured by any observer along a timelike curve, 
\begin{equation}
    \left(T_{\mu\nu}-\frac{T}{2}g_{\mu\nu}\right)t^\mu t^\nu \ge 0.
\end{equation}
The DEC states that the energy-momentum flux measured by an observer must be causal and directed forward along the observer's proper time. In the last case, the NEC represents a variation of the WEC, in which the timelike vector is replaced by a null vector $k^\mu$,
\begin{equation}
    T_{\mu\nu}k^\mu k^\nu \ge 0.
\end{equation}
These conditions are interconnected through a causal relationship, wherein DEC implies WEC, which in turn implies NEC. Among the four conditions, NEC is regarded as the weakest.

The equations for these conditions involve the energy-momentum tensor; however, they can also be expressed in terms of the Ricci and Einstein tensors. For a perfect fluid, the energy-momentum tensor is:
\begin{equation}
    T^{\mu\nu} = (\rho+P)u^\mu u^\nu + Pg^{\mu \nu},
\end{equation}
where $\rho$ represents the energy density, $P$ denotes the pressure, and $u^\mu$ indicates the 4-velocity of the fluid, which satisfies $u^\mu u_\mu = -1$.

To obtain $\rho$ and the pressures, we need to solve the Einstein Field Equations modified to incorporate the ASG parameter \eqref{eq:Einstein_ASG}. Using \eqref{eq:grr} and \eqref{eq:gthetatheta} we can express $P_r$ and $P_t$ as follows:
\begin{equation}\label{eqpr}
P_r = \frac{1}{8\pi}\Bigg[2\left(1-\frac{S}{r}\right)\,\frac{\Phi'}{r} -\frac{S}{r^3}\Bigg]\left[1+\left(\frac{\xi}{r}\right)^2\right].
\end{equation}
\begin{equation}\label{eqpt}
    P_t = \frac{1}{8\pi}\left(1-\frac{S}{r}\right)\left[\Phi''+(\Phi')^{2}+\frac{\left(S-rS'\right)}{2r(r-S)}\Phi'+\frac{\left(S-rS'\right)}{2r^2(r-S)}+\frac{\Phi'}{r}\right]\left[1+\left(\frac{\xi}{r}\right)^2\right],
\end{equation}
with the last term inside the brackets in \eqref{eqpr} and \eqref{eqpt} representing the contribution from the ASG.

In summary, the energy conditions can be expressed as functions of $\rho$ and $P$:
\begin{enumerate}
    \item \textbf{NEC}: $(P_i+\rho)\ge 0$.
    \item \textbf{WEC}: $\rho \ge 0$, $(P_i + \rho) \ge 0$.
    \item \textbf{DEC}: $(\rho - |P_i|)\ge 0$.
    \item \textbf{SEC}: $(\rho + \sum\limits_{i=1}^3 P_i)\ge 0$, $(P_i+\rho)\ge 0$.
\end{enumerate}

\begin{figure}[!htp]
\centering
\includegraphics[width=1\textwidth]{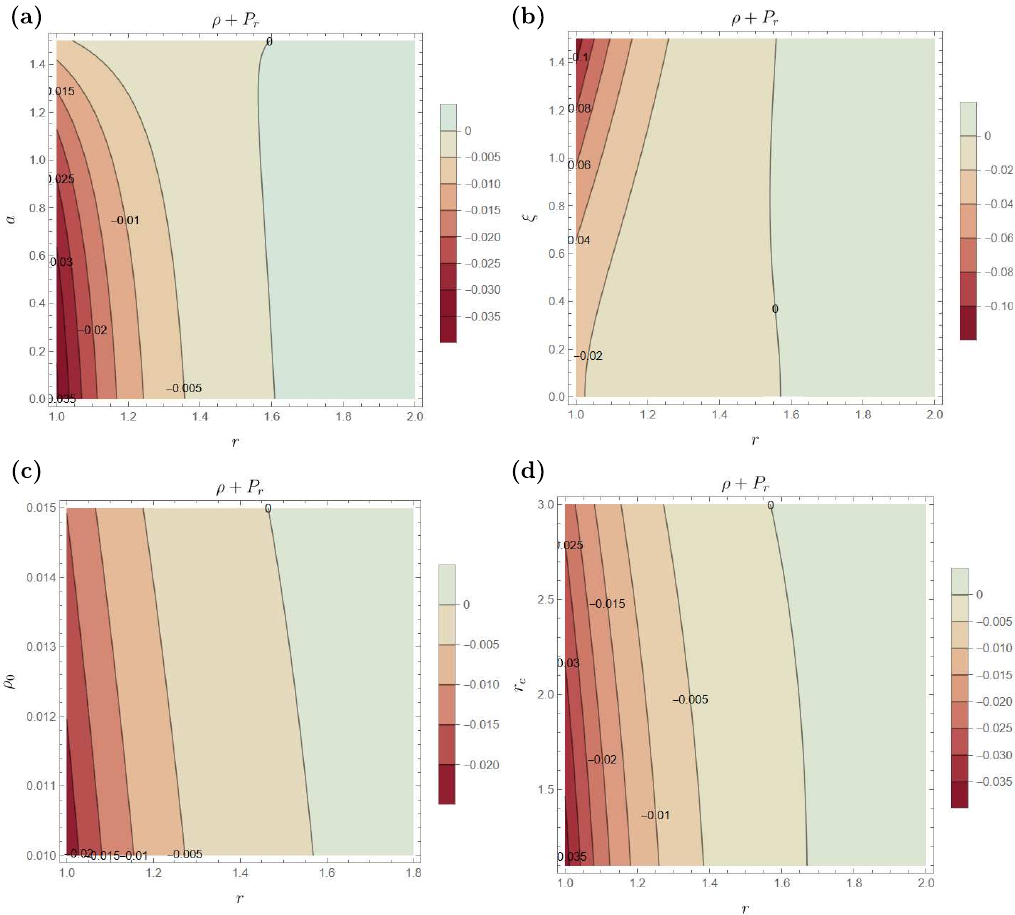}
\caption{The illustrations of radial NEC, $\rho+P_r$, for different keys parameters as a function of $r$, with: (a) $\rho_0 = 0.01, r_c = 3.0$ and $\xi =0.1$. (b) $\rho_0 = 0.01, a = 1.0$ and $r_c = 3.0$. (c) $ a = 1.0, r_c = 3.0$ and $\xi =0.1$. (d) $\rho_0 = 0.01, a = 1.0$ and $\xi =0.1$. In all panels $r_0 = 1.0$.}\label{fig5}
\end{figure}

Since we have the expressions for $\rho$ and $P$, we can analyze the behavior of the energy conditions at the throat of the wormhole. For NEC, the weakest condition, the radial component is represented by $P_r+\rho \ge 0$. Considering \eqref{eq:g00} and \eqref{eqpr}, we derive the radial NEC equation for any $r$ as
\begin{equation}\label{pemaisrho}
    P_r+\rho = \frac{1}{8\pi}\left[2\left(1-\frac{S}{r}\right)\frac{\Phi'}{r}-\frac{S}{r^3}+\frac{S'}{r^2}\right]\left[1+\left(\frac{\xi}{r_0}\right)^2\right].
\end{equation}
At the throat, $r=r_0$, we have
\begin{equation}\label{necneg}
    8\pi(P_r+\rho)\rvert_{r=r_0} = \left[\frac{S'(r_0)-1}{r_0^2}\right]\left[1+\left(\frac{\xi}{r_0}\right)^2\right].
\end{equation}
Since the throat geometric condition defines $S(r_0)=r_0$, the first term of Eq. (\ref{pemaisrho})  vanishes. Thus, the NEC signal at the throat is independent of $\Phi(r_0)$. Furthermore, the terms that do not vanish represent the ``flare-out'' condition at the throat ($S'(r_0)-1\le 0$), and, as previously mentioned, this relation is negative.  Given that the ASG contribution is always positive, we find the NEC to be consistently negative, as illustrated in Fig. \ref{fig5} (i.e., the radial NEC condition is violated). This implies that incorporating $G(r)$ directly into the Einstein field equations increases the amount of exotic matter at the throat. Therefore, it is expected that when we increase $\xi$, the exoticity of the wormhole will also increase. Additionally, \eqref{necneg} underscores the fundamental importance of the redshift function for ensuring a positive NEC beyond the throat. If $\Phi(r)$ is zero or constant, the NEC will be negative for $r>r_0$. This underscores the significance of the redshift function concerning the amount of exotic matter necessary to create a traversable wormhole, thereby preventing the need for an infinite quantity of exotic matter \cite{Garattini2019}. In summary, incorporating $G(r)$ directly into Einstein's equations cannot change the type of matter at the throat. This ASG approach does not alter the essence of Einstein's equations, as demonstrated when including $G(r)$ directly in the Einstein-Hilbert action or metric \cite{Nilton2021,Nilton2022}.

\begin{figure}[!htp]
\centering
\includegraphics[width=1\textwidth]{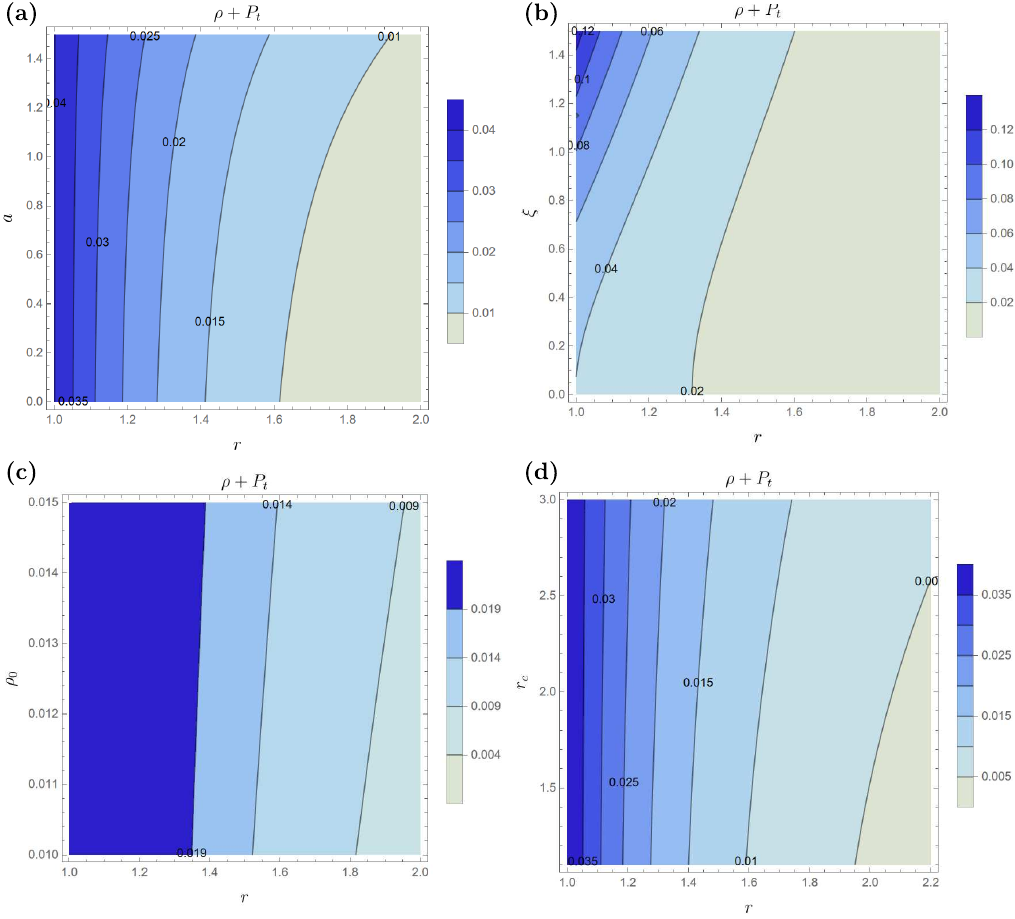}
\caption{The illustrations of tangential NEC, $\rho+P_t$, for different keys parameters as a function of $r$, with: (a) $\rho_0 = 0.01, r_c = 3.0$ and $\xi =0.1$. (b) $\rho_0 = 0.01, a = 1.0$ and $r_c = 3.0$. (c) $ a = 1.0, r_c = 3.0$ and $\xi =0.1$. (d) $\rho_0 = 0.01, a = 1.0$ and $\xi =0.1$. In all panels $r_0 = 1.0$.}\label{fig6}.
\end{figure}

The tangential NEC at the throat exhibits behavior that is opposite to that of the radial component. We can express the tangential NEC at the throat as
\begin{equation}\label{nectanpos}
    8\pi(P_t+\rho)\rvert_{r=r_0} = -\frac{1}{2r_0^2}\left[S'(r_0)-1\right]\left[\Phi'(r_0)+1\right]\left[1+\left(\frac{\xi}{r_0}\right)^2\right].
\end{equation}
Since then $S'(r_0)-1 \le0$, the $\Phi'(r_0)>0$, for construction, and the ASG term is positive definite, the tangential NEC at the throat is not violated, as shown in Fig.\ref{fig6}. In contrast to the radial component, the $\xi$ parameter enhances the positivity of this condition for $r=r_0$. The violation of the radial NEC, while the tangential NEC is upheld, signifies the presence of anisotropic exotic matter, which is essential for keeping the structure traversable. The radial NEC violation is a critical feature that generates a strong negative pressure or tension along the direction passing through the wormhole's throat. This creates a repulsive gravitational effect that maintains the throat open against collapse. Concurrently, the satisfaction of the tangential NEC indicates that the matter behaves non-exotically in the directions surrounding the throat. This fundamental difference between radial and tangential pressures defines the matter as anisotropic, effectively localizing the anti-gravitational behavior only to the axis where it is structurally necessary. This arrangement is often considered a more physically efficient and plausible configuration for a stable wormhole \cite{Jusufi2019, Mustafa2023,Garattini2019,Rahaman2016}.

\begin{figure}[!htp]
\centering
\includegraphics[width=1\textwidth]{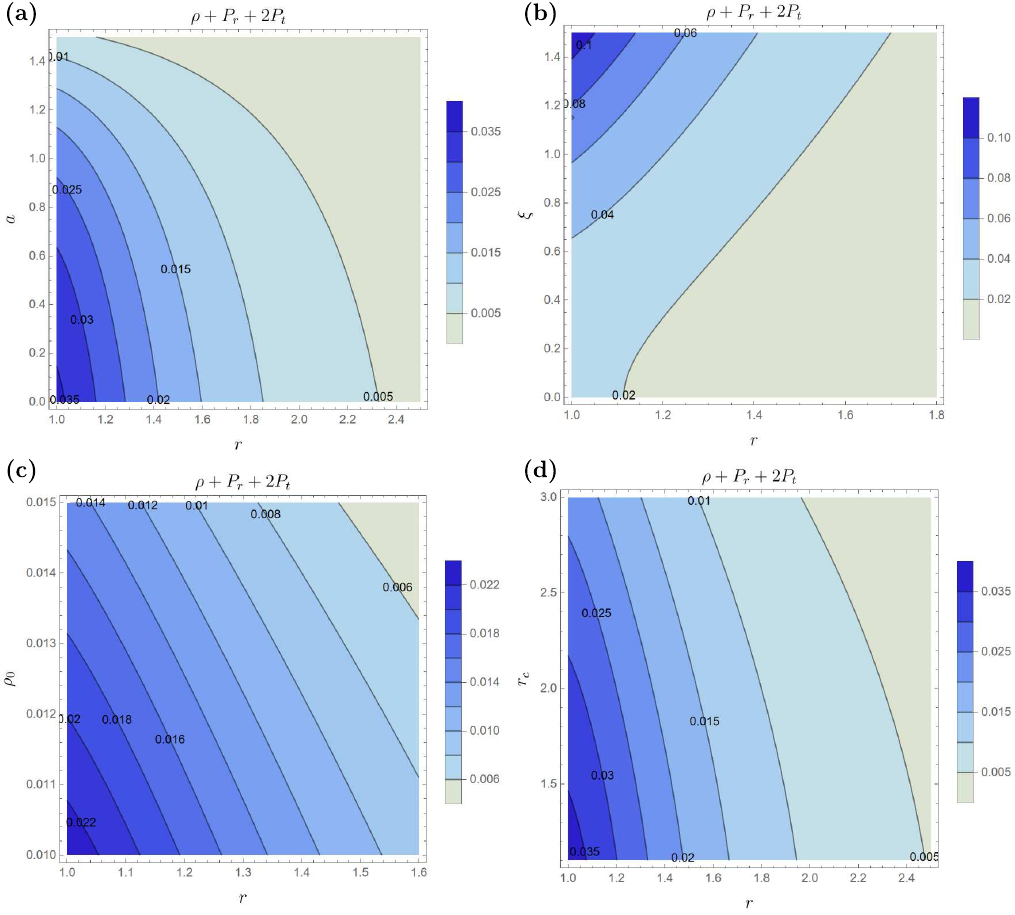}
\caption{The illustrations of SEC, $\rho+P_r+2P_t$, for different keys parameters as a function of $r$, with: (a) $\rho_0 = 0.01, r_c = 3.0$ and $\xi =0.1$. (b) $\rho_0 = 0.01, a = 1.0$ and $r_c = 3.0$. (c) $ a = 1.0, r_c = 3.0$ and $\xi =0.1$. (d) $\rho_0 = 0.01, a = 1.0$ and $\xi =0.1$. In all panels $r_0 = 1.0$.}\label{fig7}
\end{figure}

\begin{figure}[!htp]
\centering
\includegraphics[width=1\textwidth]{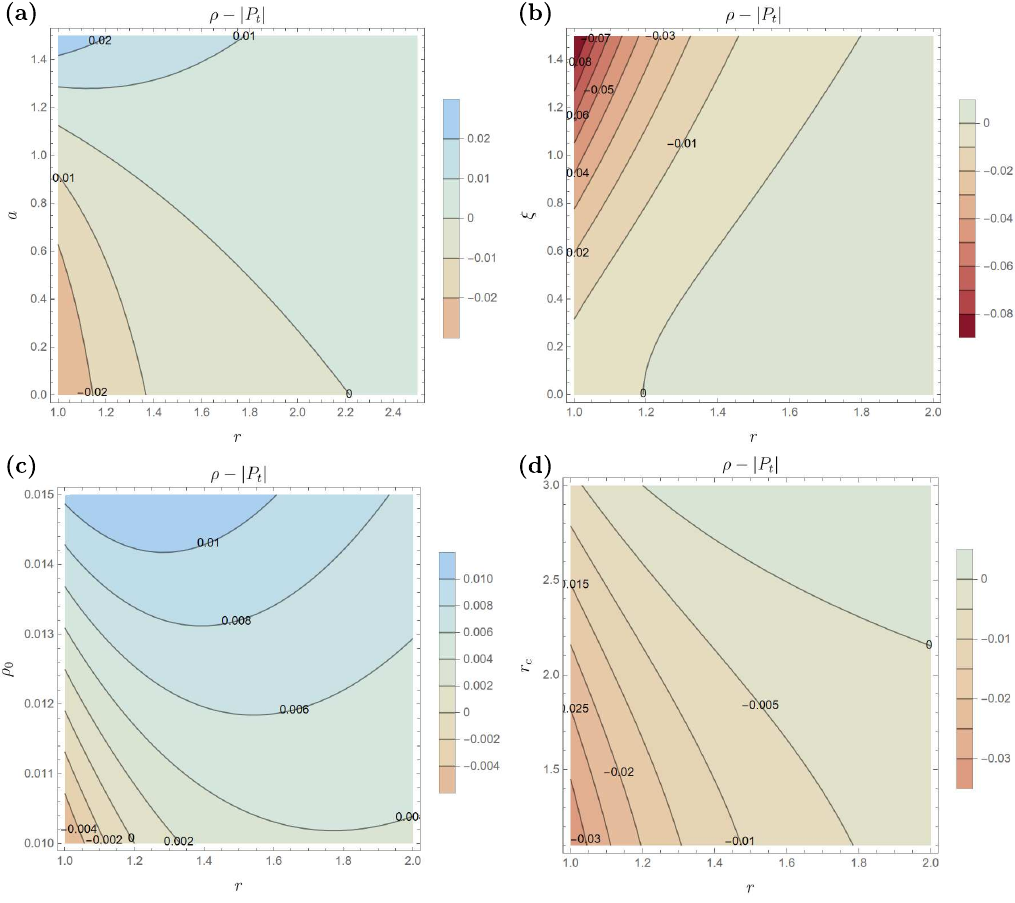}
\caption{The illustrations of tangential DEC, $\rho-|P_t|$, for different keys parameters as a function of $r$, with: (a) $\rho_0 = 0.01, r_c = 3.0$ and $\xi =0.1$. (b) $\rho_0 = 0.01, a = 1.0$ and $r_c = 3.0$. (c) $ a = 1.0, r_c = 3.0$ and $\xi =0.1$. (d) $\rho_0 = 0.01, a = 1.0$ and $\xi =0.1$. In all panels $r_0 = 1.0$.}\label{figdec}
\end{figure}

The analysis of energy conditions is illustrated in several plots shown in Figs. \ref{fig5}, \ref{fig6}, \ref{fig7} and \ref{figdec}. The radial and tangential NEC were examined within a valid range of parameters. Additionally, we analyzed the SEC and the tangential DEC. Other energy conditions were excluded from the plots, as they were considered redundant for our purposes. 

Figure \ref{fig5} provides a critical analysis of NEC, represented by $\rho + P_r$. As previously mentioned, the violation of this condition is essential for the existence of a traversable wormhole, as it indicates the presence of exotic matter. The panels clearly demonstrate that the structural parameters of the dark matter halo play a decisive role in the viability of the wormhole. Panel (a) reveals an inverse relationship between the inner slope parameter, $a$, and the extent of the NEC violation. Profiles with a large $a$ tend to suppress the exotic region, confining it to a smaller area near the throat and rendering the violation less pronounced. This indicates that a high central concentration of dark matter is detrimental to the structure of the wormhole. This suppressive effect is also consistently observed for the other halo parameters. Panel (c) shows that an increase in the characteristic density $\rho_0$ systematically weakens the NEC violation. Similarly, panel (d) illustrates that a larger characteristic radius $r_c$, which corresponds to a more spatially extended halo, also diminishes the region where $\rho+P_r < 0$. These results strongly suggest that more massive and extended dark matter distributions are less conducive to hosting traversable wormholes in this model. In contrast, the ASG-coupled parameter $\xi$ exhibits the opposite behavior. Panel (b) shows that a larger value of $\xi$ significantly enhances the NEC violation. Therefore, while the intrinsic properties of the dark matter halo act to obstruct the geometric conditions required for traversable wormholes, the ASG theory within this framework actively supports their existence by effectively generating the necessary exotic matter region.

Conversely, the analysis of the tangential NEC reveals a contrasting behavior that is equally vital for the physical viability of the wormhole solution. For a stable structure, it's desirable that this condition be satisfied. The Fig.\ref{fig6} clearly shows that this requirement is robustly met across the entire parameter space. Panel (b) demonstrates that the ASG coupling consistently reinforces this condition.On the other hand, panels (a), (c), and (d) show that while the value of $\rho + P_t$ remains positive, it does not increase significantly with variations of the structural parameters of the dark matter halo. This dynamic is fundamental, as it suggests that the gravitational modification from the ASG theory, rather than the halo's properties, is the primary mechanism that guarantees the necessary pressure anisotropy for a viable traversable wormhole.

Figure \ref{fig7} analysis SEC, represented by the expression $\rho + P_r+2P_t$. This analysis provides a comprehensive synthesis of the matter content of the wormhole, effectively integrating the behaviors observed for both the radial and tangential NEC. As depicted across all four panels in the figure, the SEC is partially satisfied throughout the entire parameter space, since the sum of the energy density and pressures is positive in all regions. This result highlights the dominance of the tangential pressure component in the overall energy-momentum budget. This dominance is most vividly illustrated in panel (b), where the ASG coupling parameter exhibits a strong positive correlation with the SEC. Just as $\xi$ was the primary driver reinforcing the tangential NEC, its influence is amplified here by the $2P_t$ term, leading to a dramatic increase in the SEC's magnitude and confirming that the primary role of the ASG theory is to generate immense tangential pressure. Conversely, the influence of the radial pressure is revealed through the SEC's clear sensitivity to the structural parameters of the dark matter halo, a trait absent in the tangential NEC analysis. However, $P_r$ acts to weaken the SEC. The influence of this weakening effect is directly modulated by the structural parameters of the dark matter halo. This is most evident in panel (c), where an increase in the $\rho_0$ leads to a lower SEC value, indicating that a denser halo amplifies the negative, suppressive contribution of the radial pressure. The same behavior is observed in panels (a) and (d).

Figure \ref{figdec} presents a comprehensive examination of tangential DEC, expressed as $\rho - |P_t|$. The radial DEC was not plotted because it $P_r$ is always negative near the throat, rendering this plot redundant.The tangential DEC serves as a crucial diagnostic of the physical viability and stability of the wormhole configuration within the ASG-inspired dark matter halo framework. The panels collectively reveal that the tangential DEC exhibits a highly structured dependence on both the halo parameters and the ASG coupling. Panel (a) demonstrates that variations in the inner slope parameter $a$ exert a relatively mild yet systematic influence on the condition. Smaller values of $a$ lead to slight violations of the tangential DEC near the throat, while higher $a$ values help to maintain $\rho - |P_t| > 0$ across a broader radial domain, suggesting that more concentrated halo profiles tend to reinforce the local energy dominance in the tangential direction. Conversely, panel (b) reveals that $\xi$ has a more pronounced effect: increasing $\xi$ systematically suppresses the tangential DEC. This behavior highlights the intrinsic tension between the ASG modifications and classical energy conditions—while the ASG coupling promotes exoticity in the radial sector (as seen in the NEC analysis), it simultaneously enhances the tangential stresses beyond the threshold permitted by the DEC, emphasizing the anisotropic nature of the matter distribution. Panels (c) and (d) further elaborate on this interplay by examining the dependence on the characteristic density $\rho_0$ and the core radius $r_c$. A higher $\rho_0$ raises  $\rho - |P_t|$, mitigating any violation and favoring the satisfaction of the tangential DEC, which indicates that denser halos strengthen the matter's dominance over tangential pressure. Taken together, these results reveal a consistent pattern: while the ASG coupling drives the system toward anisotropic configurations with stronger tangential pressures, the intrinsic halo parameters act as stabilizing agents that counterbalance this trend.

\subsection{Stability and equilibrium conditions - adiabatic sound speed}

We now turn to the analysis of the stability of the obtained wormhole solutions by examining the average squared adiabatic sound speed of the fluid \cite{Capozziello:2022zoz}. This quantity, denoted by $\langle v_s^2 \rangle$, is defined as
\begin{equation}
  \label{vsound}
    \langle v_s^2 \rangle = \frac{d\langle P\rangle}{d\rho}
    = \frac{d\langle P\rangle/dr}{d\rho/dr},
\end{equation}
where $\langle P\rangle = (P_r + 2P_t)/3$ represents the average pressure across the three spatial directions.  
For stability, the condition $\langle v_s^2 \rangle \geq 0$ must hold, while the physical requirement $\langle v_s^2 \rangle < 1$ ensures causality is preserved.
\begin{figure}[!htp]
\centering
\includegraphics[width=0.49\textwidth]{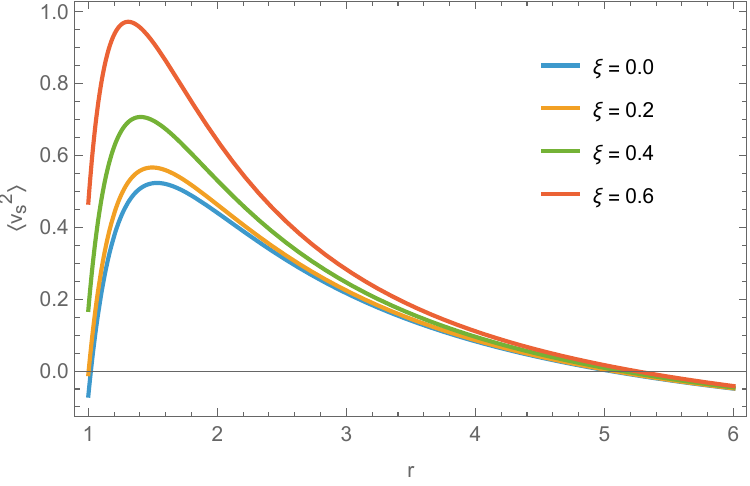}
\includegraphics[width=0.49\textwidth]{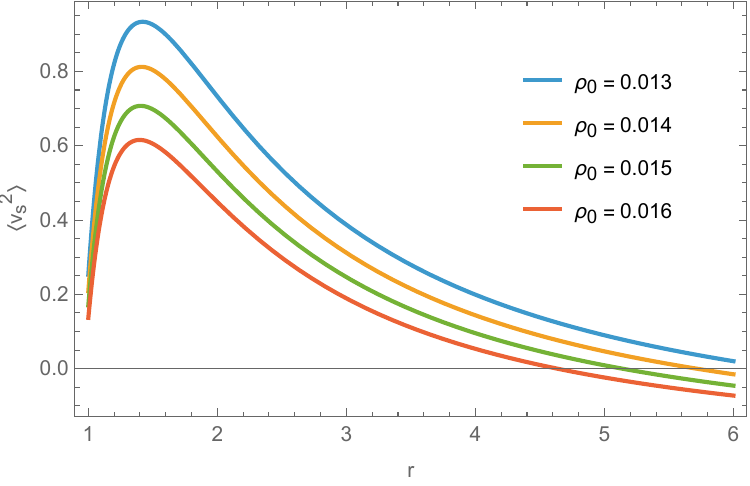}
\includegraphics[width=0.49\textwidth]{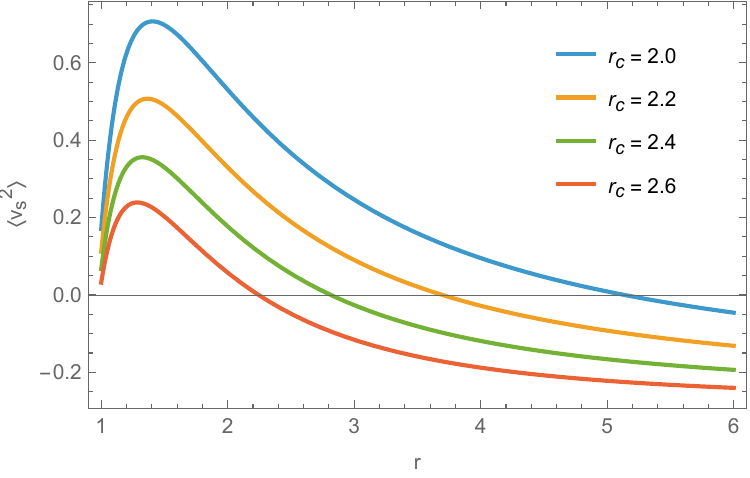}
\includegraphics[width=0.49\textwidth]{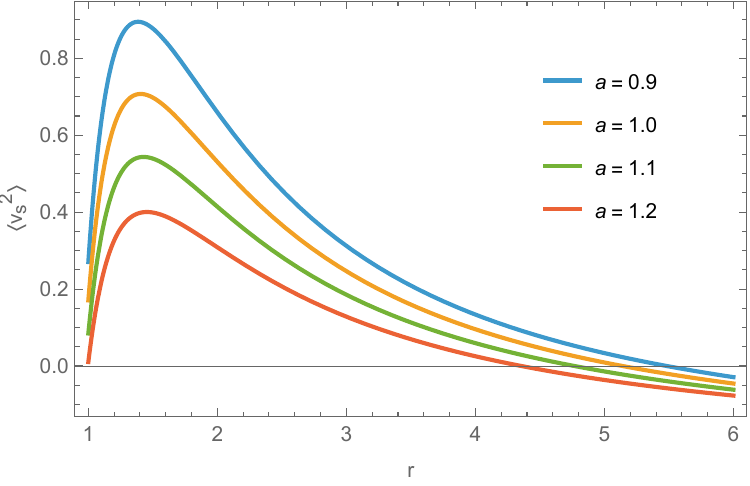}
\caption{Average squared adiabatic sound speed curves of a wormhole as a function of the radial coordinate, produced for Dekel-Zhao's dark matter density profile in the ASG framework. Upper-left  panel: Varying ASG parameter $\xi$, with $a=1.0, r_c= 2.0, \rho_0 = 0.015$. Upper-right panel: Varying dark matter density $\rho_0$, with $a = 1.0, r_c = 2.0, \xi = 0.4$. Bottom-left panel: Varying dark matter reach $r_c$, for a=1.0, $\rho_0 = 0.015, \xi = 0.4$. Bottom-right panel: Varying dark matter model according to the parameter $a$, with $r_c=2.0, \rho_0 = 0.015, \xi = 0.4$. In all panels we do $r_0 = 1.0$.}\label{fig8}
\end{figure}
FIG. \ref{fig8} depicts the behavior of the average squared adiabatic sound speed of the fluid for different parameter choices. Only a narrow range of these parameters ensures both stability and causality at the throat and in its immediate vicinity. 

As shown in the upper-left panel, higher values of the ASG parameter $\xi$ shift the profiles of $\langle v_s^2 \rangle$ upward at and near the throat, thus promoting greater extension of the allowed stability region of the system. Note that in the realm of general relativity ($\xi=0.0$), there is no stability at the throat. In the upper-right  panel, we can see that increasing central density $\rho_0$ concentrates more dark matter in the vicinity of the throat, reducing $\langle v_s^2 \rangle$ and eventually leading to negative values that indicate instability. Similarly, increasing the scale radius $r_c$ (bottom-left panel), which corresponds to a larger spatial extent of dark matter, also worsens the stability, as it increases the overall influence of dark matter on the fluid structure. Finally, the bottom-right panel illustrates the role of the parameter $a$, which characterizes a deformation of the NFW dark matter profile ($a=1$ recovers the standard case). Larger values of $a$ further spoil the stability by decreasing $v_s^2$ at and near the throat (since the deformation steepens the inner density slope, thereby enhancing pressure gradients around the throat). These results confirm thus that both a higher concentration and a larger extent of dark matter spoil the stability of the fluid, whereas the ASG parameter acts in the opposite direction, favoring the maintenance of a stable configuration.


\subsection{Stability and equilibrium conditions - modified TOV equation}

In the framework of ASG, the running of the gravitational coupling \( G(r) \) modifies the local conservation law of the energy–momentum tensor. Instead of the usual condition \( \nabla_{\mu} T^{\mu}_{\nu} = 0 \), one obtains
\begin{equation}
\nabla_{\mu} \big[G(r) T^{\mu}_{\nu}\big]
= \big[\nabla_{\mu} G(r)\big] T^{\mu}_{\nu}
+ G(r)\, \nabla_{\mu} T^{\mu}_{\nu}=0,
\end{equation}
which expresses the coupling between the scale dependence of gravity and the matter sector. This generalized conservation law directly leads to a modified Tolman–Oppenheimer–Volkoff (TOV) equation that governs the equilibrium of anisotropic configurations. The resulting equation reads
\begin{equation}
-\,G'(r)\,P_{r}(r) 
+ G(r)\!\left[-\big(\rho(r) + P_{r}(r)\big)\Phi'(r) 
- P_{r}'(r) 
+ \frac{2}{r}\big(P_{t}(r) - P_{r}(r)\big)\right] = 0.
\end{equation}
This equation can be interpreted as the balance among three main forces: The gravitational force \( \mathcal{F}_{g} = -G(r)\,(\rho + P_{r})\,\Phi'(r) \), which acts inward; the hydrostatic force \( \mathcal{F}_{h} = -G(r)\,P_{r}'(r) \), which opposes the gravitational pull; and the anisotropic force \( \mathcal{F}_{a} = 2G(r)(P_{t} - P_{r})/r \), which arises when the radial and tangential pressures differ. The additional term proportional to \( G'(r) \) introduces a new contribution of quantum nature, namely
\begin{equation}
\mathcal{F}_{\mathrm{Q}} = -\,G'(r)\,P_{r}(r),
\end{equation}
which may be interpreted as a \emph{scale-dependent gravitational force} induced by the spatial variation of the coupling. This new force embodies the quantum correction predicted by ASG, acting as an effective source of disequilibrium that can either enhance or counterbalance the classical gravitational attraction depending on the sign of \( G'(r) \). Consequently, the overall equilibrium of the system near the wormhole throat becomes sensitive to the running behavior of the gravitational constant.

\begin{figure}
    \centering
    \includegraphics[width=0.49\linewidth]{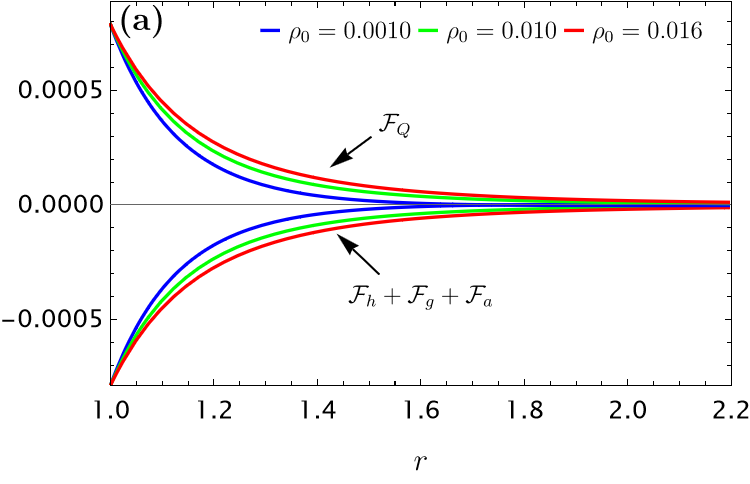}
    \includegraphics[width=0.49\linewidth]{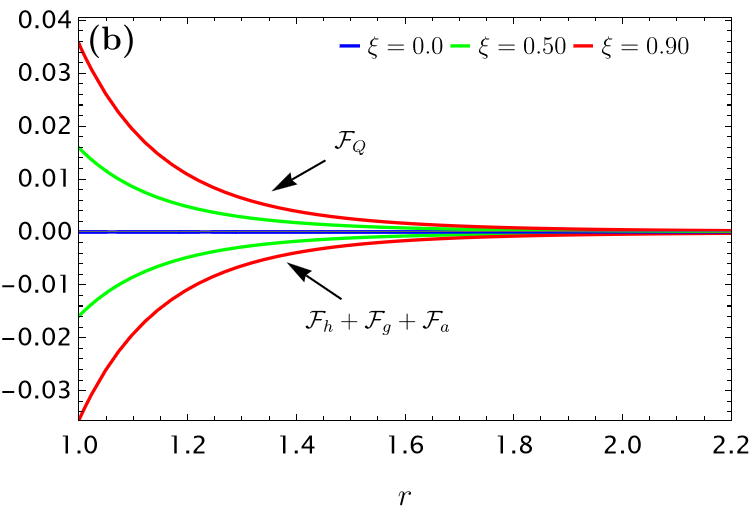}
    \includegraphics[width=0.49\linewidth]{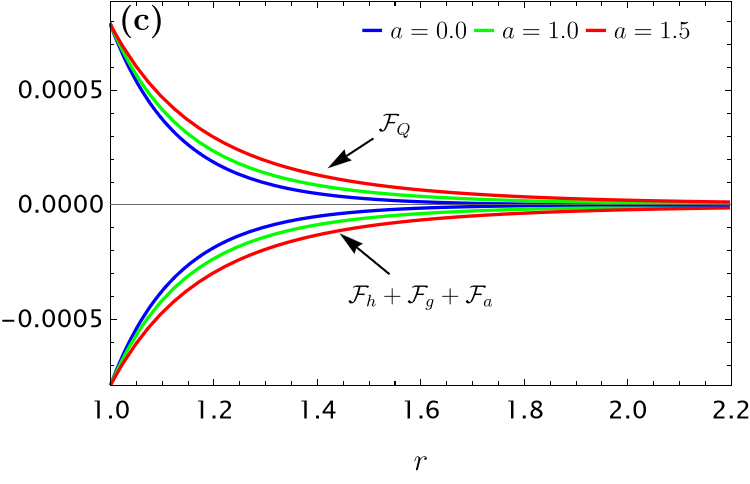}
    \includegraphics[width=0.49\linewidth]{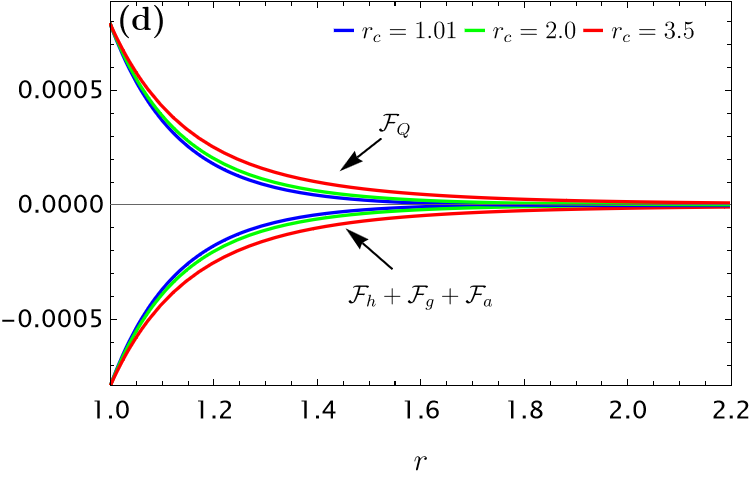}
    \caption{Profiles of the combined hydrostatic ($\mathcal{F}_h$), gravitational ($\mathcal{F}_g$), and anisotropic ($\mathcal{F}_a$) forces in contrast with balance force ($\mathcal{F}_Q$), shown as a function of the radial coordinate $r$. The panels illustrate the dependency on different model parameters: (a) $a=1.0, \, r_c = 3.0$ and $\xi = 0.1$. (b) $a=1.0, \, r_c = 3.0$ and $\rho_0 = 0.01$. (c) $\rho_0=0.01, \, r_c = 3.0$ and $\xi = 0.1$. (d) $a=1.0, \, \rho_0 = 0.01$ and $\xi = 0.1$. In all panels $r_0 = 1.0$.}
    \label{forces}
\end{figure}

As shown in Fig. \ref{forces}, the inclusion of $\mathcal{F}_{Q}$ is essential to restore equilibrium in the system, since in its absence the total force does not vanish throughout the space. The magnitude and sign of $\mathcal{F}_{Q}$ act to counterbalance the excess of the gravitational pull, effectively stabilizing the configuration against collapse. The plots indicate that $\mathcal{F}_{Q}$ behaves as a repulsive component, increasing with the strength of the ASG corrections and providing an outward pressure that compensates the inward gravitational force. Therefore, in this framework, the equilibrium condition $\mathcal{F}_{g}+\mathcal{F}_{h}+\mathcal{F}_{a}+\mathcal{F}_{Q}=0$ defines a new balance of forces, ensuring that the wormhole structure remains sustained even when the effective gravitational coupling varies with the distance scale. Moreover, the intensity of these balanced forces becomes more pronounced not only with increasing $\xi$, but also for higher values of the characteristic dark matter density $\rho_{0}$ and its spatial extension $r_{c}$, as well as with deviations of the Zhao parameter $a$ from the NFW limit ($a=1$), revealing the joint influence of both ASG and the dark matter halo on the equilibrium of the wormhole. In the classical limit, where $G(r)\to G_{0}=1$ with $\xi$ vanishing, the standard equilibrium condition of General Relativity is naturally recovered.

\section{Phenomenological features}\label{secPheF}

While analyses of black hole shadows are phenomenologically more relevant for spinning geometries, we focus here on the static configuration to isolate and systematically evaluate the impact of ASG corrections on the shadow morphology. This simplified approach represents a crucial first step toward confronting the theoretical model developed in this work with actual observational data \cite{Liu2019pov, Kimet2021, Jiang_2024, LIU2024139052}. Our investigation is particularly timely given the remarkable progress in recent studies that utilize shadow properties to constrain parameters of alternative and extended gravity theories \cite{Okyay_2022, PANTIG2023169197, CIMDIKER2021100900, Vagnozzi:2022moj}.

In this section, we provide an estimation of the shadow radius that enables us to establish preliminary bounds on the free parameters of our model. For computational tractability and to isolate the fundamental geometric effects, we neglect potential influences from accretion flows and assume negligible variations due to observer position and inclination. We consider a distant observer situated in the asymptotic region, for which the shadow radius is given by:
\begin{equation}  
    R_{sh} \approx R_o \sin{\alpha_{sh}},  
\end{equation}  
where \( R_o \) denotes the observer's position \cite{PERLICK20221}. The angular radius of the shadow \( \alpha_{sh} \) follows the expression:  
\begin{equation}  
\sin{\alpha_{sh}} = \frac{\gamma(r_{ph})}{\gamma(R_o)},  
\end{equation}  
with the metric function defined as:  
\begin{equation}  \label{gamma}
\gamma(r) = \sqrt{-\frac{g^{tt}}{g^{\phi\phi}}}. 
\end{equation}  
The critical photon sphere radius \( r_{ph} \) is determined by solving the orbital condition:  
\begin{equation}  \label{r_ph}
    \frac{d\gamma^2(r)}{dr} = 0 \quad \text{at} \quad r = r_{ph}.  
\end{equation}  

From Eqs. (\ref{redshift}) and (\ref{gamma}), we find that Eq. (\ref{r_ph}) yields
\begin{equation}
   \frac{2(r_{ph}-r_0) (r_{ph}+r_0)^3}{r_{ph}^3}=0,
\end{equation}
and therefore $r_{ph}$ coincides with $r_0$, which significantly simplifies the analysis while maintaining the essential physical characteristics. 

Another key aspect of our approach involves the identification of the running energy scale of ASG with the local spacetime curvature:
\begin{equation}\label{prescription}
\xi\sim\frac{1}{\sqrt{R(r_0)}},
\end{equation}
where the Ricci scalar is evaluated at the wormhole throat \( r_0 \), resulting in
\begin{eqnarray}
\xi^{-2}=-\frac{2 \left( \frac{r_0}{r_c} \right)^{-a}}{r_0^2 (r_0 + r_c)^3 (r_0^2 + \xi^2)}
\Bigg[ 
r_0^5 \left( \frac{r_0}{r_c} \right)^{a}
+ 3 r_0^4 r_c \Bigg( \left( \frac{r_0}{r_c} \right)^{a} 
- 8 \rho_0 \pi r_c^2 \left( \frac{r_0 + r_c}{r_c} \right)^{a} \Bigg) 
+ 3 r_0 \left( \frac{r_0}{r_c} \right)^{a} r_c^2 \xi^2
+ \left( \frac{r_0}{r_c} \right)^{a} r_c^3 \xi^2
 & & \nonumber \\ 
+ r_0^3 \left( \frac{r_0}{r_c} \right)^{a} \left( 3 r_c^2 + \xi^2 \right)
+ r_0^2 \left( \frac{r_0}{r_c} \right)^{a} r_c \left( r_c^2 + 3 \xi^2 \right)
\Bigg]. \nonumber\\
\end{eqnarray}
This identification establishes an implicit dependence of the throat radius $r_{0}$ on the ASG parameter $\xi$, and hence of the shadow radius as well. Variations in $\xi$ therefore directly modulate both the geometric scale of the wormhole and the corresponding shadow size.
\begin{figure}[!htp]
\centering
\includegraphics[width=0.6\textwidth]{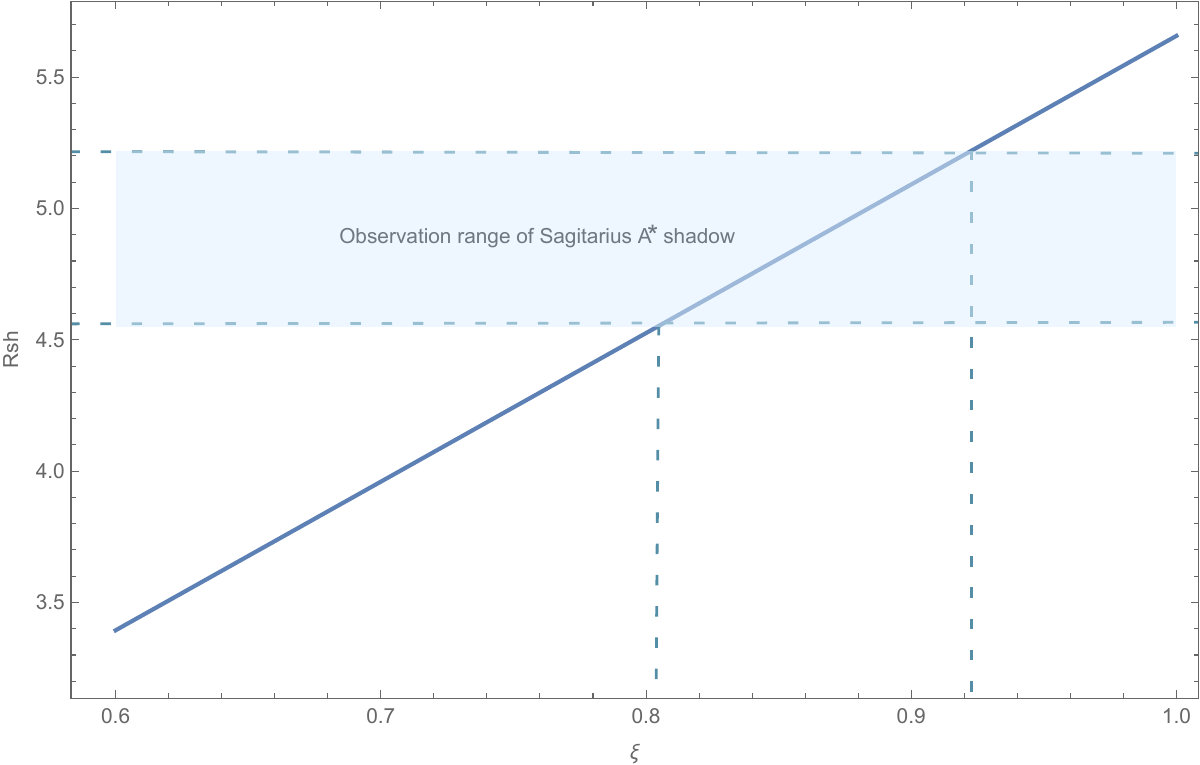}
\caption{Shadow radius $R_{\mathrm{sh}}/M$ of the wormhole sourced by dark matter described by the Dekel-Zhao density profile (parameter $a=1$) within the ASG framework, shown as a function of the parameter $\xi$. The shaded horizontal band marks the Event Horizon Telescope (EHT) observational range for Sgr A*. The comparison illustrates the range of $\xi/M$ values for which the wormhole shadow becomes consistent with the angular size measured by the EHT.}
\label{fig.shadow}
\end{figure}
\begin{figure}[!htp]
\centering
\includegraphics[width=0.5\textwidth]{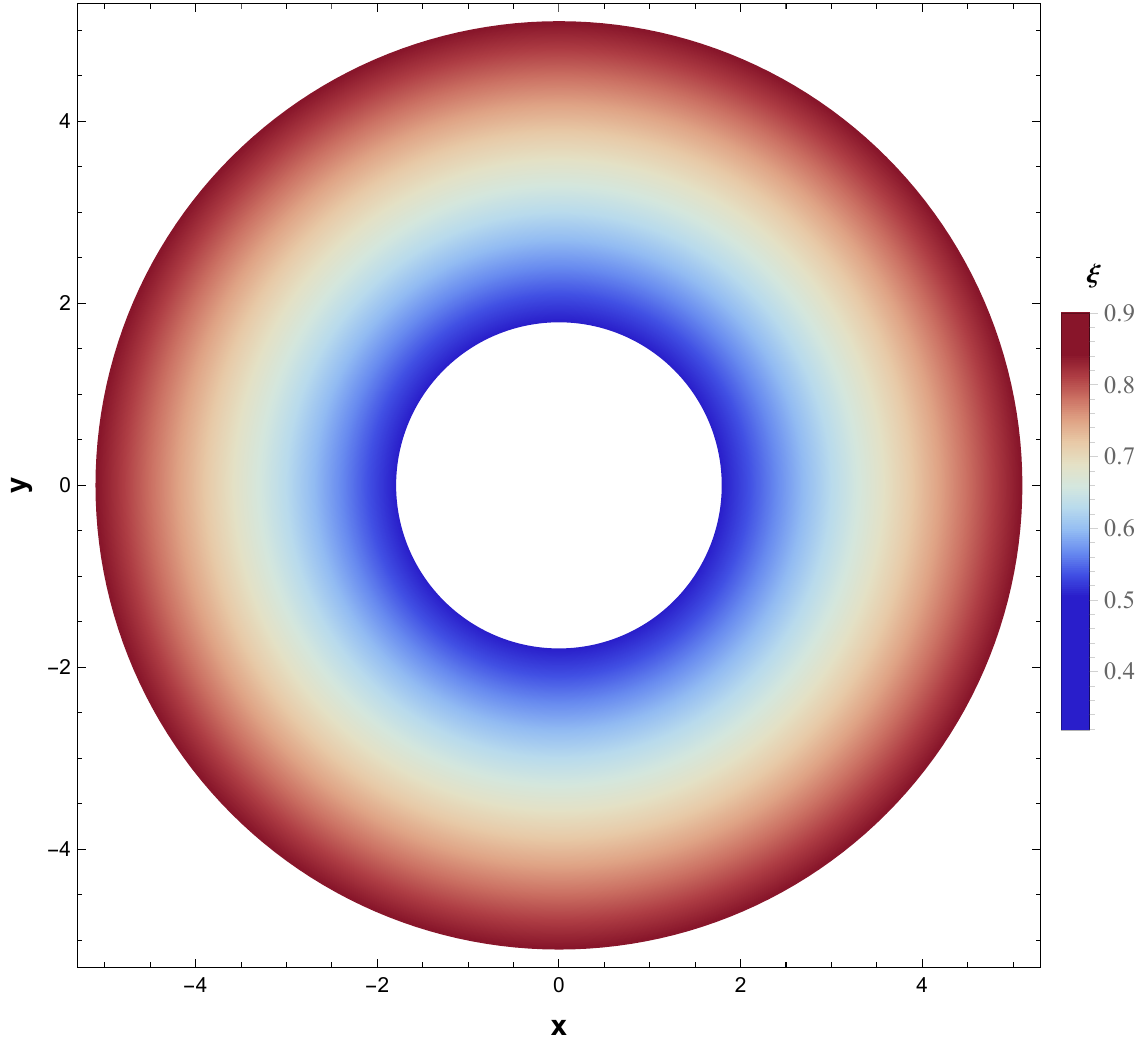}
\caption{Visual representation of the wormhole shadow in ASG with Dekel-Zhao dark matter ($a = 1$). The color gradient encodes the variation of the parameter $\xi$ from $0.3$ (blue, inner shadows) to $0.9$ (red, outer shadows), clearly demonstrating the monotonic increase of the shadow radius $R_{\text{sh}}$ with $\xi$. This visual pattern provides an intuitive understanding of how the ASG parameter influences the observable shadow characteristics.}
\label{fig.shadow2}
\end{figure}

Fig.~\ref{fig.shadow} presents thus the computed shadow radius for the NFW dark matter model, \(a = 1\), employing the characteristic parameters \(r_{c} = 8~\mathrm{kpc}\) and \(\rho_{0} = 0.5~\mathrm{GeV/cm^{3}}\). In geometric units \((G = c = 1)\), these correspond to \(r_{c} \simeq 4.18 \times 10^{10} M\) and \(\rho_{0} \simeq 2.3 \times 10^{-29} M^{-2}\). The analysis reveals that the theoretical prediction for the shadow radius \(R_{\mathrm{sh}}/M\) increases approximately linearly with the ASG parameter \(\xi\) over the explored range of values. When compared with the observational constraints established by the Event Horizon Telescope (EHT) for Sgr~A*, \(R_{\mathrm{sh}}/M \in [4.55, 5.22]\), we find a significant overlap between the theoretical curve and the EHT observational band for \(\xi/M\) values in the range \(\sim 0.80\)–\(\sim 0.92\). Interestingly, the lower bound of $\xi$ found in this analysis is comparable to the upper bound reported in \cite{Kumar:2019ohr} for the Schwarzschild black hole also in the IR ASG framework. This close agreement indicates that ASG-corrected wormholes sourced by a Dekel-Zhao dark matter halo can successfully reproduce the apparent shadow size observed at the Galactic center, offering a potentially observable signature of such exotic compact objects. At the same time, this also highlights the difficult task of differentiating between wormholes and black holes based solely on their shadow appearance. Further analysis of possible gravitational wave signatures or some specific lensing effects might offer another path to distinguish between different compact objetcs surrounded by dark matter halos.

The visual representation in Figure \ref{fig.shadow2} further elucidates the relationship between the ASG parameter $\xi$ and the shadow morphology. The sequence of circular curves demonstrates a clear positive correlation: as $\xi/M$ increases from $0.3$ to $0.9$, the shadow radius $R_{\text{sh}}/M$ progressively expands, transitioning from compact blue inner shadows to more extensive red outer shadows. The wormhole shadow radius increases with the ASG parameter because the gravitational coupling weakens, reducing the effective gravitational attraction and thereby enlarging the size of photon orbits around the wormhole throat. This direct dependence between the quantum-gravity scale parameter and the observable shadow size provides a crucial observational signature that may facilitate the detection and characterization of such exotic objects within modified gravity scenarios.

\section{conclusions}\label{seccon}

In this work, we have obtained novel solutions for traversable wormholes within the framework of Asymptotically Safe Gravity (ASG), sourced by a galactic dark matter halo described by the Dekel-Zhao density profile, and investigated their properties. Our approach explicitly incorporated the scale dependence of the gravitational constant $G$ in the field equations, as prescribed by the renormalization group flow of ASG in the infrared (IR) regime, neglecting, therefore, backreaction effects. This formulation allowed for a consistent treatment of quantum gravitational corrections at galactic scales, where $G$ effectively runs with the energy scale. The analysis revealed that the interplay between ASG corrections, parameterized by the running scale $\xi$, and the dark matter distribution, characterized by the inner slope $a$, characteristic density $\rho_{0}$, and scale radius $r_{c}$, fundamentally shapes the wormhole’s geometry and determines its physical viability.

Geometric analyses showed that scale-dependent gravitational coupling $G(r)$, arising from ASG, together with dark matter, plays a decisive role in shaping the geometry of the wormhole. The modified shape function satisfies the flare-out and asymptotic flatness conditions only within restricted domains of the ASG and dark matter parameters, indicating that traversability demands a fine balance between quantum corrections and halo characteristics. The curvature study also showed that increasing the ASG parameter $\xi$ enhances the curvature concentration around the throat, signaling a stronger local geometric distortion induced by the running of the gravitational coupling. This effect is clearly corroborated by the embedding diagrams, which display a sharper throat and a more pronounced spatial curvature as $\xi$ grows. Altogether, these results demonstrate that ASG corrections intensify the local geometric structure of the wormhole, while the dark matter distribution governs the global extension and smoothness of the spacetime, establishing a coherent and physically consistent configuration.

The physical analysis of the ASG wormhole supported by the dark matter halo initially confirmed that the Null Energy Condition (NEC) is necessarily violated at the throat and, consequently, the remaining energy conditions, a typical signature of traversable wormholes. Such a violation becomes more pronounced for higher values of the ASG parameter $\xi$, as the running of the gravitational coupling intensifies the curvature concentration in that region. In contrast, larger characteristic dark matter densities $\rho_{0}$ and greater halo extents $r_{c}$ tend to alleviate violation, as the enhanced matter contribution partially compensates for the exotic energy required in the throat. The physically acceptable redshift function further mitigates this effect away from the throat, reducing the overall amount of exotic matter needed to sustain the configuration. However, the SEC is only partially violated across the entire domain, since the sum of energy density and pressures remains positive in all regions. Moreover, an increase in quantum effects through the parameter $\xi$ tends to enhance the fulfillment of this condition, whereas a higher concentration or spatial extension of dark matter acts in the opposite direction.

The study of the equilibrium through the average squared sound speed further revealed the competing roles of dark matter and quantum corrections: higher central densities $\rho_{0}$, larger scale radii $r_{c}$, and steeper inner slopes $a$ tend to destabilize the configuration by lowering $\langle v_s^2 \rangle$, whereas increasing the ASG parameter $\xi$ exerts a stabilizing influence, at least around the throat. This behavior reflects the dual nature of the ASG framework, in which the running of the gravitational coupling can simultaneously intensify curvature at the throat while enhancing dynamical stability there. 

Regarding the equilibrium conditions derived from the modified TOV equation, they were found to be consistent with the stability analysis based on the adiabatic sound speed, indicating that the wormhole remains stable only when the ASG correction compensates for the destabilizing influence of dark matter. In this sense, the additional term of quantum force $\mathcal{F}_{Q} = -G'(r)P_{r}$ balances gravitational, hydrostatic, and anisotropic forces, whose magnitudes increase with higher values of $\xi$, $\rho_{0}$, $r_{c}$, and with deviations of the Zhao parameter $a$ from the NFW limit. The sensitivity of the equilibrium configuration to variations in $\xi$ is particularly pronounced, emphasizing the dynamical role of the ASG coupling scale. Overall, these results show that the quantum gravitational effects within the ASG not only modify the geometric structure of the wormhole but also play a decisive role in ensuring its physical consistency and stability.

Finally, from a phenomenological standpoint, the analysis of the shadow properties revealed a clear and testable imprint of the ASG corrections on observable scales. For the NFW dark matter profile ($a = 1$), the computed shadow radius exhibited an almost linear growth with the ASG parameter $\xi$ in the analyzed interval, reflecting the direct influence of the running gravitational coupling on the photon trajectories near the wormhole throat. This dependence is physically justified by relating the ASG scale with the Ricci's curvature evaluated at the throat. This prescription naturally links the quantum-gravity scale to the region of maximal curvature, allowing us to infer how variations in $\xi$ modulate the shadow radius. Remarkably, within the range $\xi/M \sim 0.80$--$0.92$, the predicted shadow size agrees with the observational bounds reported by the Event Horizon Telescope (EHT) for Sgr~A*. This concordance indicates that the interplay between ASG corrections and the Dekel-Zhao dark matter halo can reproduce realistic shadow dimensions without invoking rotation or additional free parameters. Moreover, the monotonic expansion of the shadow with increasing $\xi$, confirmed by the polar visualizations, suggests that the quantum-gravity scale leaves a distinct and potentially measurable signature in the strong-field regime. Altogether, these results emphasize the phenomenological relevance of ASG-corrected wormholes as viable candidates for exotic compact objects and provide a concrete pathway to probe the infrared behavior of quantum gravity through precision shadow observations of galactic centers.

\section*{Acknowledgments}
CRM would like to thank Conselho Nacional de Desenvolvimento Cient\'{i}fico e Tecnol\'ogico (CNPq) for the partial financial support, through grant 301122/2025-3. FBL is funded by Fundação Cearense de Apoio ao Desenvolvimento Científico e Tecnológico (FUNCAP) and by  Conselho Nacional de Desenvolvimento Científico e Tecnológico (CNPq), grant number 305947/2024-9. EO acknowledges support from FUNCAP(BP6-0241-00335.01.00/25).

\bibliographystyle{apsrev4-1}
\bibliography{ref.bib}
\end{document}